\documentclass[11pt]{article}
\pdfoutput=1
\usepackage{jpub,amsmath,amssymb,hyperref}
\usepackage{bm}
\usepackage{amsmath}
\usepackage{amssymb}
\usepackage{youngtab}
\usepackage{bbold}
\usepackage{pgfplots,todonotes}
\usepackage{aas_macros}
\usepackage{tikz}
\usepackage[latin1]{inputenc}
\usepackage{breqn}
\usetikzlibrary{arrows,positioning,decorations.markings,decorations.pathmorphing,calc}
\definecolor{hyperref}{RGB}{026,028,087}
\usepackage{placeins}
\usepackage{tikz-feynman}

\def\gsim{ \lower .75ex \hbox{$\sim$} \llap{\raise .27ex \hbox{$>$}} }
\def\lsim{ \lower .75ex \hbox{$\sim$} \llap{\raise .27ex \hbox{$<$}} }
\def\be{\begin{equation}}
\def\ee{\end{equation}}
\def\bea{\begin{eqnarray}}
\def\eea{\end{eqnarray}}

\def\L{{\cal L}}
\def\mpl{M_{\rm Pl}}
\newcommand{\mn}{\mu\nu}

\newcommand{\commentout}[1]{}

\newcommand{\MPl}{M_{\rm Pl}}

\newcommand{\comment}[1]{}

\newcommand{\bs}{\begin{split}}

\newcommand{\G}{  \bar{G} }
\newcommand{\aB}{\alpha_B}
\newcommand{\aM}{\alpha_M}
\newcommand{\aT}{\alpha_T}
\newcommand{\aK}{\alpha_K}
\newcommand{\aI}{\alpha_i}

\newcommand{\smi}[1]{{\textcolor{black}{#1}}}

\def\d{\mathrm{d}}
\def\({\left(}
\def\){\right)}

\newcommand*{\mathcolor}{}
\def\mathcolor#1#{\mathcoloraux{#1}}
\newcommand*{\mathcoloraux}[3]{%
\protect\leavevmode
\begingroup
\color#1{#2}#3%
\endgroup
}
\newlength{\stheight}
\newcommand\textst[1][fu-grey]{
\ifmmode\setlength{\stheight}{+1.0ex}
\else\setlength{\stheight}{+0.5ex}
\fi
\bgroup\markoverwith{\textcolor{#1}{\rule[\the\stheight]{2pt}{1.0pt}}}\ULon
} 
\newcommand{\textins}[2][fu-grey]{
\ifmmode\mathcolor{#1}{#2}
\else\textcolor{#1}{#2}\@\,
\fi
}
\graphicspath{{./}}
\allowdisplaybreaks

\usepackage{titlesec}
\titleformat{\part}{\Large\bfseries}{}{0pt}{Part \thepart\ --\ }

\tikzstyle{vecArrow} = [thick, decoration={markings,mark=at position
1 with {\arrow[semithick]{open triangle 60}}},
double distance=1.4pt, shorten >= 5.5pt,
preaction = {decorate},
postaction = {draw,line width=1.4pt, white,shorten >= 4.5pt}]

\begin{document}

\title{Positivity Bounds on Dark Energy: \\ When Matter Matters}

\author[a,b]{Claudia de Rham,}
\author[c]{Scott Melville,}
\author[c,d]{Johannes Noller}

\affiliation[a]{Theoretical Physics, Blackett Laboratory, Imperial College London, SW7 2BZ, U.K.}
\affiliation[b]{CERCA, Department of Physics, Case Western Reserve University, 10900 Euclid Ave, Cleveland, OH 44106, USA}
\affiliation[c]{DAMTP, University of Cambridge, Wilberforce Road, Cambridge, CB3 0WA, U.K.}
\affiliation[d]{Institute of Cosmology \& Gravitation, University of Portsmouth, Portsmouth, PO1 3FX, U.K.}

\emailAdd{c.de-rham@imperial.ac.uk}
\emailAdd{scott.melville@damtp.cam.ac.uk}
\emailAdd{johannes.noller@port.ac.uk}

\date{today}
\abstract{
Positivity bounds---constraints on any low-energy effective field theory imposed by the fundamental axioms of unitarity, causality and locality in the UV---have recently been used to constrain scalar-tensor theories of dark energy. However, the coupling to matter fields has so far played a limited role. We show that demanding positivity when including interactions with standard matter fields leads to further constraints on the dark energy parameter space. We demonstrate how implementing these bounds as theoretical priors affects cosmological parameter constraints and explicitly illustrate the impact on a specific Effective Field Theory for dark energy. We also show in this model that the existence of a standard UV completion requires that gravitational waves must travel superluminally on cosmological backgrounds.
}

\keywords{Positivity Bounds, Effective Field Theory, Dark Energy}

\setcounter{tocdepth}{2}
\maketitle

\section{Introduction}

Challenging our understanding of the Universe and  General Relativity (GR) is a central goal of modern cosmology. Despite its many successes, GR cannot be the fundamental description of our Universe---for instance, it is only an effective description of gravity at low energies (breaking down at least at the Planck scale, if not below). In parallel,  accounting for the late-time acceleration of the Universe leads to the well-known cosmological constant problem.
GR may therefore require modifications on both theoretical and phenomenological grounds.
Fortunately, in recent years there has been significant progress in developing model-independent parameterised approaches that allow for a systematic exploration of dark energy/modified gravity effects in a (linear) cosmological setting \cite{Gubitosi:2012hu,Bloomfield:2012ff,Gleyzes:2014rba,Bellini:2014fua,Gleyzes:2013ooa,Kase:2014cwa,DeFelice:2015isa,Langlois:2017mxy,Frusciante:2019xia,Renevey:2020tvr,Lagos:2016wyv,Lagos:2017hdr}, resulting in a variety of cosmological parameter constraints on deviations from GR from (current and forecasted) experimental data, see e.g. \cite{Noller:2018wyv,BelliniParam,Hu:2013twa,Raveri:2014cka,Gleyzes:2015rua,Kreisch:2017uet,Zumalacarregui:2016pph,Alonso:2016suf,Arai:2017hxj,Frusciante:2018jzw,Reischke:2018ooh,Mancini:2018qtb,Brando:2019xbv,Arjona:2019rfn,Raveri:2019mxg,Perenon:2019dpc,SpurioMancini:2019rxy,Bonilla:2019mbm,Baker:2020apq,Joudaki:2020shz,Noller:2020lav,Noller:2020afd}.

As with any effective field theory (EFT), these parameterised approaches remain agnostic about the nature of the underlying UV completion. While this greatly improves efficiency (allowing the translation of observations into model-independent constraints), it introduces the risk that certain regions of parameter space may be secretly unphysical or ``unstandard''---what could seem a perfectly consistent EFT may not have any healthy UV (high-energy) completion, or may not enjoy any local UV completion compatible with standard axioms. To ensure that the underlying UV theory respects fundamental properties---such as unitarity, causality and locality---the low-energy EFT must satisfy various constraints, known as ``positivity bounds''. Following many recent advances in these EFT bounds and their consequences for dark energy and modified gravity \cite{Adams:2006sv,Jenkins:2006ia,  Adams:2008hp, Nicolis:2009qm, Bellazzini:2014waa, Bellazzini:2015cra, Baumann:2015nta,  Cheung:2016yqr,Bonifacio:2016wcb,deRham:2017avq,deRham:2017zjm,deRham:2018qqo,Bellazzini:2016xrt,deRham:2017imi,deRham:2017xox}, it is now more important than ever to incorporate these constraints when performing precision tests for cosmology and for physics beyond GR.

In a cosmological setting, one often separates the matter fields from the gravitational sector and may even model them completely separately (e.g. as a perfect fluid, rather than as dynamical degrees of freedom). However, since all fields couple gravitationally, they ultimately mediate scattering processes which must obey positivity arguments if the EFT is to have a standard (unitary, causal, local) UV completion. This applies to light, baryonic matter as well as any other matter field living in a dark or decoupled sector, irrespectively of whether or not those fields are relevant to the dynamics or phenomenology of the EFT in question.
We will therefore state the following very explicitly,
\begin{quote}

A low-energy EFT has no standard UV completion if it violates the positivity bounds for scattering between \emph{any} of its low-energy degrees of freedom.

\end{quote}
This observation is widely appreciated, but we emphasise it nonetheless since it strictly strengthens the power of positivity constraints on any given EFT. Rather than applying bounds to the scattering of one particular species (or a small number of species), the previous statement is making explicit the potential constraining power in scattering all possible combinations of all possible fields---including, in the case of cosmology, scattering quantized matter fields with each other and with fields in the gravitational sector.
In some situations, accounting for these additional positivity bounds can lead to constraints that are orthogonal to what was otherwise considered as common wisdom (eg. see \cite{deRham:2019ctd,deRham:2020zyh,Alberte:2020jsk}).

At this stage it may be worth commenting on the notion of ``standard" UV completion which is implicit in the applicability of the positivity bounds. By standard UV completion, we have in mind the EFT to be the low-energy limit in the Wilsonian sense of a local, unitary, Lorentz invariant and causal weakly-coupled high-energy completion in which the Froissart bound is satisfied. Note that the assumption of weak coupling does not require a tree-level completion as is sometimes further implicitly assumed in the literature. We refer to the violation of any of these assumptions as a ``non-standard" UV completion, \cite{Dvali:2010jz,Dvali:2010ns,Dvali:2011nj,Dvali:2011th,Vikman:2012bx,Kovner:2012yi,Keltner:2015xda}.
Such non-standard UV completions may either be non-weakly coupled, or may for instance include a small violation of locality or micro-causality.

\paragraph{Gravitational scalar field theory:}
Throughout we will be working within the context of a gravitational EFT that contains a light scalar degree of freedom $\phi$ \cite{Gleyzes:2013ooa} that may for instance play the role of dark energy.
Moreover, we shall, for simplicity, restrict ourselves to a shift-symmetric Horndeski theory \cite{Horndeski:1974wa,Deffayet:2009wt}, also known as Weakly Broken Galileons \cite{Pirtskhalava:2015nla}. Specifically, we will focus on
\begin{align}\label{quarticA}
\hspace{-0.3cm}S = \int \d^4x \sqrt{-g} \Big\{ \Lambda_2^4 G_2(X) + \MPl^2 G_{4}(X) R + \Lambda_2^4 G_{4,X}(X) \left( [\Phi]^2-[\Phi^2] \right) +\L_{\rm matter}(\psi, g) \Big\}\,,
\end{align}
where the dimensionless matrix $ \Phi_{\mu}^{\; \nu} \equiv   \nabla_\mu \nabla^\nu\phi / \Lambda_3^3$ (with $[ \Phi^n]$ denoting the trace, e.g. $[\Phi^2]\equiv \nabla^\mu\nabla_\nu\phi\nabla^\nu\nabla_\mu\phi / \Lambda_3^6$), $G_2$ and $G_4$ are functions of the dimensionless $X \equiv -\tfrac{1}{2 \Lambda_2^4} g^{\mu\nu}\partial_\mu\phi\partial_\nu\phi$, and $\MPl \gg \Lambda_2 \gg \Lambda_3$ are constant scales which characterise the EFT.
$\L_{\rm matter}(\psi, g)$ indicates the Lagrangian for all the matter fields $\psi$ which, in this theory, are assumed to be minimally coupled to the metric $g_\mn$. Introducing non-minimal couplings between a matter field and the metric $g_\mn$ or with the scalar field $\phi$ would further affect the EFT.
The power counting in \eqref{quarticA} ensures that the operators appearing at the scale $\Lambda_3$ are protected by Galileon invariance, and although this is broken by gravitational corrections, since graviton exchange is suppressed by at least one factor of $\MPl$ this ensures that a hierarchy $\Lambda_2^4 \sim \MPl \Lambda_3^3$ is radiatively stable (see \cite{Luty:2003vm,Nicolis:2004qq,deRham:2010eu,Burrage:2010cu,deRham:2014wfa})\footnote{Note that the radiatively stable nature of such theories can be maintained in the presence of at least some specific sets of shift-symmetry-breaking interactions -- see \cite{deRham:2012ew,Noller:2018eht,Heisenberg:2020cyi} for examples and a more detailed discussion of this point.}.

The scalar-tensor theory \eqref{quarticA} corresponds to a particular shift-symmetric subset of the Horndeski scalar-tensor theory \cite{Horndeski:1974wa,Deffayet:2009wt}, in which cubic and quintic interactions have been turned off.
This is the same example theory previously explored in Ref.~\cite{Melville:2019wyy} and has the particularly nice feature that positivity constraints are easily mapped onto constraints on the effective parameters controlling linearised cosmological perturbations \cite{Bellini:2014fua}. Here we expand on the analysis of \cite{Melville:2019wyy}, where positivity bounds from the scattering of dark energy scalars were used to constrain cosmological parameters, by deriving and applying additional positivity bounds that arise in the presence of matter degrees of freedom. 
\smi{In particular, we focus on the subset of Horndeski theories for which the functions $G_2 (X)$ and $G_4 (X)$ obey the following two conditions:}
\begin{itemize}

\item[(i)] \smi{in addition to a cosmological solution for $\phi$, there is also a stable Minkowski solution $\phi = 0, \, g_{\mu\nu} = \eta_{\mu\nu}$, }

\item[(ii)] \smi{the effective couplings $G_{4,X}$ and $G_{4,XX}$ do not receive large corrections from the cosmological background so are comparable in both backgrounds.}

\end{itemize}
\smi{For this subset of scalar-tensor theories, Lorentz-invariant positivity bounds around the stable Minkowski solution can be imported to the cosmological solution and compared with data.  
While there are interesting theories which may violate one or both of the above assumptions, by focussing on this particular subset we are able to demonstrate explicitly that positivity constraints from the UV can have important consequences for how we analyse data. }

\paragraph{Constraints from speed of gravitational waves:} Following the direct detection of gravitational waves from the Neutron star merger GW170817 with optical counterpart, \cite{TheLIGOScientific:2017qsa,Monitor:2017mdv,GBM:2017lvd}, the speed of gravitational waves at LIGO frequencies is constrained to be luminal within one part in $10^{15}$. Trusting EFTs of dark energy at order $10^2$Hz would then lead to ruling out any model for which the speed of gravitational waves differs from unity, including the model considered in  \eqref{quarticA} \cite{Lombriser:2015sxa,Lombriser:2016yzn,Creminelli:2017sry, Sakstein:2017xjx, Ezquiaga:2017ekz, Baker:2017hug, Akrami:2018yjz, Heisenberg:2017qka, BeltranJimenez:2018ymu}. Note however that since the theory breaks down at or below the cutoff $\Lambda_3\sim 10^2$Hz, we do not expect that EFT to be meaningful on those scales \cite{deRham:2018red} and in what follows we contemplate the possibility that \eqref{quarticA}  remains an acceptable low-energy EFT at sufficiently low energies relevant for dark energy and the late-time acceleration of the Universe.

\paragraph{Positivity bounds:}
Unitarity implies the positivity of the coefficients of the partial wave expansion of the elastic $2-2$ amplitude $\mathcal{A}$, between two massive particles on a flat background.  The simplest  bounds use the positivity of the first coefficients and the $s\leftrightarrow u$ crossing symmetry, \cite{Adams:2006sv} (see also \cite{Pham:1985cr,Ananthanarayan:1994hf,Pennington:1994kc,Comellas:1995hq} or earlier discussion of this constraint in chiral perturbation theory),  assuming a causal (analytic in energy), and local (polynomially bounded growth at high energies) UV completion places constrains on the Wilson coefficients appearing in $\mathcal{A}$ (see also \cite{Jenkins:2006ia,Dvali:2012zc}).

There is, however, an infinite number of bounds that can be derived from the requirement of unitarity \cite{deRham:2017avq,deRham:2017zjm,deRham:2018qqo}, and all bounds can further be improved by appropriate substraction of the light loops contributions up to the cutoff of the EFT \cite{Bellazzini:2016xrt,deRham:2017imi,deRham:2017xox}. Lorentz invariance however implies full $s\leftrightarrow u \leftrightarrow t$ crossing symmetry and this information was seldom used until recently. Indeed full crossing symmetry was recently implemented directly at the level of the positivity bounds
 \cite{Bellazzini:2020cot,Tolley:2020gtv,Caron-Huot:2020cmc,Sinha:2020win}, where it was shown to further constrain massive Galileons \cite{deRham:2017imi} and other scalar field theories with weakly broken symmetries.
 Typically, the direct implementation of these bounds to the gravitational context is challenging, most notably due to the presence of a $t$-channel pole and concrete models are known to slightly violate the positivity bounds in the gravitational setup \cite{Alberte:2020jsk,Alberte:2020bdz}, with a resolution provided in \cite{Caron-Huot:2021rmr}. In this work, we shall be working in a decoupling limit $\mpl\to \infty$ where issues related to the $t$-channel pole may be evaded. Moreover, we will focus on one of the simplest Lorentz-invariant bounds from unitarity and $s \leftrightarrow u$ crossing symmetry. This will be sufficient to illustrate our main point: that the coupling to matter inevitably generates additional bounds from dark energy-matter scattering.

The positivity bounds we shall consider will require Lorentz invariance. We shall therefore  first consider the scattering of small fluctuations about the trivial Minkowski background $\phi = 0$, which ensures a Lorentz-invariant scattering amplitude and then import the resulting constraints to a cosmological background, invoking the covariant nature of \eqref{quarticA}.
Such an approach is justified when assuming that the cosmological solutions we consider here smoothly connect with a trivial Minkowski vacuum.  One could in principle go further, in particular there has been recent developments in establishing positivity bounds directly to Lorentz-breaking backgrounds \cite{Grall:2021xxm}, but we leave these considerations for further studies.

In the case we shall be interested in, when expanded in powers of the center of mass energy, $s$, and the momentum transfer, $t$ and taking the $\mpl\to \infty$, the amplitude takes the form
\begin{equation}
\mathcal{A} (s,t) =  c_{ss} ~ \frac{s^2}{\Lambda_2^4}  + c_{sst} ~ \frac{ s^2 t}{\Lambda_3^6} +  ... \; ,
\label{eqn:Ast}
\end{equation}
and UV requirements then demand that $c_{ss} > 0$.
Going beyond the forward limit, we will use positivity in the form,
\begin{align}
 c_{sst} \geq - \frac{3 \Lambda_3^4}{2 \Lambda_2^4} c_{ss}\,,
 \label{eqn:csst_bound}
\end{align}
first given in \cite{deRham:2017avq, deRham:2017zjm} (see also \cite{Pennington:1994kc, Vecchi:2007na, Manohar:2008tc, Nicolis:2009qm, Bellazzini:2016xrt} for earlier discussion of positivity at finite $t$), assuming\footnote{
Note that if the EFT breaks down at some low scale, $\epsilon \Lambda_3\ll \Lambda_3$, then the bound \eqref{eqn:csst_bound} would be $c_{sst} \geq - \frac{3 \Lambda_3^4}{2 \epsilon^2 \Lambda_2^4} c_{ss}$. Neglecting the small ratio $(\Lambda_3/\Lambda_4)^4 \sim ( H / \mpl)^{2/3}$ only requires that the EFT be valid at scales much above $( H / \mpl )^{1/3}  \Lambda_3 \sim H$, and so this assumption is equivalently a necessary condition for describing the expanding spacetime background.
} that the EFT can be used to subtract the contribution from light loops up to the scale $\Lambda_3$.
Notionally, these bounds on $c_{ss}$ and $c_{sst}$ are diagnosing whether it is possible (even in principle) for some new physics to enter at the scales $\Lambda_3$ and $\Lambda_2$ to restore unitarity in the full UV amplitude. If these bounds were violated, it would indicate that this new high energy physics is of a non--standard type, as indicated earlier, either due to the violation of the weakly coupled completion, violation of micro-causality or mild violation of locality \cite{Dvali:2010jz,Dvali:2010ns,Dvali:2011nj,Dvali:2011th,Vikman:2012bx,Kovner:2012yi,Keltner:2015xda}.
By itself, observing a violation of the positivity bounds would have groundbreaking consequences for our understanding of high-energy physics.

For the scalar-tensor theory \eqref{quarticA}, we show below that, in addition to the positivity bounds from scalar-scalar scattering found in \cite{Melville:2019wyy},
\begin{align}
\frac{\bar G_{4,X}}{\bar{G}_4} &\geq - 2 \bar{G}_{2,XX}  \,  , &   \frac{\bar{G}_{4,X}^2}{\bar{G}_4}  &\leq - \bar G_{4,XX}  \, ,
\label{eqn:ScaPos}
\end{align}
there are separate bounds from scalar-matter scattering,
\begin{align}
 \bar G_{4,X}  \geq  0  \; ,
 \label{eqn:MatPos}
\end{align}
where the $\bar{G}_n$ are the functions evaluated on the Minkowski background $\phi = 0$. 
Interestingly, the new bound \eqref{eqn:MatPos} places a qualitatively different restriction on the low-energy EFT, and in particular requires that the speed of tensor modes (to which  matter fields -- including light -- couple minimally) is strictly \emph{greater} than the speed of any matter field (including photons)---gravitational waves are superluminal \cite{deRham:2019ctd,deRham:2020zyh}.

When compared with observational data, we show that the bounds \eqref{eqn:ScaPos} and \eqref{eqn:MatPos} can be implemented as priors to improve parameter estimation (assuming that $G_{2,X} > 0$ so that the flat vacuum is stable).
Interestingly, we find that once the prior \eqref{eqn:ScaPos} is imposed,
the resulting observational constraints are already highly consistent with positivity in the matter sector \eqref{eqn:MatPos}.
We also show how this outcome would have been dramatically different had one instead imposed other priors that do not rely on the same assumptions of causality and unitarity.

The rest of the manuscript is organized as follows: In section~\ref{sec-posHorn}, we review the positivity bounds inferred from scalar-scalar scattering in a limit where gravity decouples and issues from the $t$-channel pole are irrelevant. We then derive the new positivity bounds inferred from scalar-matter scattering. In section~\ref{sec-cosHorn} we show how the positivity priors affect the outcome of observational constraints on the parameters of the dark energy EFT. In particular we show how positivity priors differ from standard stability and subluminality criteria. We end with a discussion and outlook in section~\ref{sec:discussion} and leave some of the technical details to Appendix~\ref{app:A}.

\section{Positivity bounds} \label{sec-posHorn}

\paragraph{Positivity of scalar-scalar scattering:}
As shown in \cite{Melville:2019wyy}, (see also Appendix~\ref{app:A}), expanding \eqref{quarticA} about a flat background ($g_{\mu \nu} = \eta_{\mu \nu} + h_{\mu \nu} /\MPl$) with zero-vev for the scalar field \smi{$(\phi = 0 + \varphi)$},
then canonically normalizing $\varphi$ such that $\bar{G}_{2,X} = 1/2$, the tree-level scattering amplitude for $\varphi \varphi \to \varphi \varphi$ takes the form \eqref{eqn:Ast}, with,
\begin{align}
  c_{sst} = - 6 \left(  \bar{G}_{4,XX} + \bar{G}_{4,X}^2  /  \G_4 \right)  \; , \;\;\;\; c_{ss} = 2 \bar{G}_{2, XX}  +   \bar{G}_{4,X}  /  \G_4\,,
\end{align}
where an overbar indicates that the function is evaluated on the flat background.
The bounds $c_{ss}>0$ and \eqref{eqn:csst_bound} therefore become,
\begin{align} \label{posbounds}
\frac{\bar G_{4,X}}{\bar{G}_4} &\geq - 2 \bar{G}_{2,XX}  \,  , &   \frac{\bar{G}_{4,X}^2}{\bar{G}_4}  &\leq - \bar G_{4,XX}  \, ,
\end{align}
where we have assumed $\Lambda_2 \gg \Lambda_3$. The other elastic amplitudes, $\varphi h \to \varphi h$ and $h h \to h h$, vanish at $\mathcal{O} (1/\mpl)$ (with $\Lambda_3$ fixed), and so scattering with external gravitons does not impose any additional constraints in the $\mpl \to \infty$ decoupling limit.
If one goes beyond this decoupling limit by including the subleading $\mathcal{O} \left( 1 / \mpl \right)$ corrections, the massless $t$-channel pole affects these bounds \cite{Alberte:2020jsk,Alberte:2020bdz,Caron-Huot:2021rmr}. Our bounds \eqref{posbounds} can nonetheless be consistently applied with the understanding that they are subject to small corrections of order $\mathcal{O} \left( \Lambda_3 / \mpl  \right)$, (such corrections are of the same order as the $\mathcal{O} \left( \Lambda_3^4 / \Lambda_2^4  \right)$ corrections in \eqref{eqn:csst_bound} which we have already neglected, see Appendix~\ref{app:A} for details).

Note that the above amplitudes have been derived on a flat background.
We will assume that these bounds are still applicable on cosmological backgrounds. This amounts to assuming that $G_{2,X}$ remains positive for fluctuations about both backgrounds, where we implicitly invoke the fully covariant nature of \eqref{quarticA} to suggest that such backgrounds can be smoothly connectable.

\begin{table}[t!]
\begin{center}
{\renewcommand{\arraystretch}{1.4}
\setlength{\tabcolsep}{0.1cm}
\begin{tabular}{|l||crcc|} \hline
{} & $c_B$ & $c_M$\qquad \quad & & $c_T$\\ \hline\hline
no priors & $0.71\substack{+0.90 \\ -0.71}$ &\qquad $-0.02\substack{+1.32 \\ -0.89}$ \qquad  & \hspace{0.2cm} & $ -1^* \le c_T < 0.25$\\ \hline
$\varphi \chi$ prior & $0.54\substack{+0.77 \\ -0.62}$ & $0.47\substack{+1.17 \\ -0.88}$ & & $0^* \leq c_T < 0.78$ \\ \hline
$\varphi\varphi$ prior & $0.26\substack{+0.46 \\ -0.46}$ & $0.67\substack{+0.97 \\ -0.58}$ & & $0.46\substack{+0.64 \\ -0.41}$\\ \hline
both priors & $0.26\substack{+0.46 \\ -0.46}$ & $0.67\substack{+0.97 \\ -0.58}$ & & $0.46\substack{+0.64 \\ -0.39}$\\ \hline
\end{tabular}
}
\end{center}
\caption{Posteriors on the dark energy/modified gravity $c_i$ parameters \eqref{oParam} for the quartic Horndeski theory \eqref{quarticA} as displayed in Fig.~\ref{fig1}, i.e. following from different combinations of positivity priors -- see \eqref{posprior} and \eqref{lumprior}. Uncertainties shown denote the $95 \%$ confidence level. The distribution for $c_T$ can be strongly skewed. We therefore do not give a mean value in such cases and denote limit values due to prior boundaries (when there is an excellent fit to the data on that boundary) with an asterisk. As expected from fig. \ref{fig1}, the constraints for the last two rows are near identical, with only a minimal modification in the lower bound for $c_T$.
}
\label{tab1}
\end{table}

\paragraph{Positivity of scalar-matter scattering:}
Having considered positivity bounds from scalar-scalar scattering, we now move on to the matter sector,
assuming a universal coupling to matter, $h_{\mu\nu} T^{\mu\nu} / 2\mpl$, where $T^{\mu\nu}$ is the stress-tensor for all  matter fields. We focus on one such field, which we call $\chi$, but emphasize that there is no implicit restriction on the nature of $\chi$. In principle $\chi$ could be designating any light Standard Model field, including light or baryonic matter, as well as any other dark sector, or other particle that may exist in our Universe.
Within the low-energy EFT, by ``light'', we typically mean a particle with a mass below $\Lambda_3$, \smi{which is typically close to $\left( M_P H_0^2 \right)^{1/3} \approx 10^{-12}$eV for dark energy}, and this includes the photon.
For the other massive particles of the Standard Model, the incoming energy $s$ would necessarily be larger than $\Lambda_3^2$ in the $\varphi \chi \to \varphi \chi$ scattering process. Trusting the calculation of this amplitude for a field of mass larger than $\Lambda_3$ may require trusting the EFT beyond its regime of validity, so in all what follows we will have in mind a field $\chi$ lighter than $\Lambda_3$, for instance the photon\footnote{Note however that the positivity bounds are not applied in the physical region but rather in the small $s$ region. So while the massive particles of the Standard Model are considered as heavy field in terms of this EFT, they may still affect the bounds of the EFT coefficients at low-energy.}.
As shown in Appendix~\ref{app:A}, the specific spin of the field $\chi$ has no impact on our conclusions. In the presence of such a matter field $\chi$, we can then derive additional positivity bounds from $\varphi \chi \to \varphi \chi$ scattering.
Canonically normalizing the field $\chi$, the amplitude takes the form \eqref{eqn:Ast} with,
\begin{align} \label{posbounds2}
 c_{ss} =  \G_{4,X}
\end{align}
and positivity requires $\G_{4,X} \geq 0$.
This bound is qualitatively different from scattering without matter, and as we shall discuss below, requires in particular that the speed of tensors is always larger than that of $\chi$ whenever considering a profile that spontaneously breaks Lorentz invariance \cite{deRham:2019ctd}.

\paragraph{Speed of gravitational waves:}
We now consider cases in which the $\phi = 0$ Minkowski solution in the EFT \eqref{quarticA} smoothly connects to a cosmological background, or in fact to any other background that spontaneously breaks Lorentz invariance  (even if very softly).
As soon as we investigate a background on which $\partial_\mu \bar \phi$ is no longer null, i.e. for which $\bar X\ne 0$, then  the speed $c_{\rm GW}$ of gravitational waves on that background is given by,
\begin{align}
\label{eq:speeds}
 \frac{ c_{\rm GW}^2 }{ c_m^2 } =
\(1 -  2 \bar{X} \frac{\bar{G}_{4,X}}{\bar G_4}\)^{-{\rm sign}\bar X}
\,,
\end{align}
where we have included the speed $c_m$ of light or of any other matter field, which in the frame we consider in \eqref{quarticA}  is minimally coupled to the metric $g_\mn$ and is hence exactly luminal. Since $\bar G_4$ ought to be positive for that background to make sense (stable tensor modes), it follows that the corrections to the speed of gravitational waves as compared to that of light is always determined by the coefficient of $\bar G_{4,X}$ (independently of the sign of $\bar X$), which is precisely the same coefficient that is bounded by the scalar-matter positivity bound. It follows that the scalar-matter positivity bounds always impose the speed of gravitational waves to be larger than that of any other field $\chi$ minimally coupled to the metric $g_\mn$ in \eqref{quarticA}.

In appendix~\ref{app:A}, we show how these results hold in more generic dark energy EFTs, including for instance the quintic Horndeski term, and is a general consequence of positivity applied to any disformal matter-coupling in the Einstein frame at leading order in derivatives.

\section{Comparison with observational constraints} \label{sec-cosHorn}

Equipped with the positivity bounds of the previous section, we are now in a position to use them as theoretical priors and to investigate the impact of such priors on cosmological constraint analyses. We will closely follow the approach of \cite{Melville:2019wyy} here, with a focus on integrating the novel positivity priors discussed above.

\paragraph{Linear cosmology:}
%
The dynamics of linear perturbations around cosmological backgrounds (following the approach of Refs. \cite{BelliniParam,Alonso:2016suf}, we will assume this to be a $\Lambda{}$CDM background) for Horndeski theories is controlled by four background functions (in addition to the Hubble scale, which controls the background expansion itself): the so-called $\alpha_i$ \cite{Bellini:2014fua}.
These are the running of the effective Planck mass $\alpha_M$, the kineticity $\alpha_K$ that contributes to the kinetic energy of scalar perturbations (effectively unconstrained by linear cosmological perturbations \cite{BelliniParam,Alonso:2016suf}, so we will omit it from the subsequent analysis), the braiding $\alpha_B$ that quantifies the strength of kinetic mixing between scalar and tensor perturbations, and the tensor speed excess $\alpha_T$, which is related to the speed of sound of tensor perturbations $c_{\rm GW}$ via $c_{\rm GW}^2 = 1 + \alpha_T$.
In terms of the model functions $G_i$, and for our specific example \eqref{quarticA}, these are given by \cite{Melville:2019wyy},
\begin{align}
\aM &=-\frac{2\dot X}{HM^2}\left(G_{4,X} + 2 X G_{4,XX}\right),
&\aB &= \frac{8X}{M^2}\left(G_{4,X}+2XG_{4,XX}\right),
&\aT &= \frac{4X}{M^2} G_{4,X} \,,
\label{alpha2}
\end{align}
where $M^2 = 2\left(G_4-2XG_{4,X}\right)$. Here it is useful to re-arrange two of the above expressions and instead write them as
\begin{align}
\aB &= 2\aT+ 16\frac{X^2}{M^2}G_{4,XX}, 
&\aM &=-\frac{1}{4}\frac{\dot X}{HX}\aB.
\label{alpha3}
\end{align}
Note that all functions in \eqref{alpha2} and \eqref{alpha3} are un-barred, since we are working on a cosmological background here (recall the bar denoted evaluation on a flat Minkowski background). Having derived the positivity bounds \eqref{posbounds} and \eqref{posbounds2} on the derivatives of $G_4 (X)$, we would now like to re-cast them in a form more directly applicable to cosmological constraint analyses. Specifically, this involves relating the above bounds to the $\alpha_i$ used in this context.

\paragraph{$\varphi\varphi$ scattering prior:}  We can now translate the positivity bounds \eqref{posbounds} and \eqref{posbounds2} into priors on the $\alpha_i$. For the bounds derived from \eqref{posbounds} this is discussed in detail in \cite{Melville:2019wyy} -- here we quickly summarise the key outcomes relevant to this section.
The $c_{ss}$ bound on $\bar{G}_{2,XX}$ is not particularly constraining at this level,
since none of the $\aI$ in \eqref{alpha2} depend on $G_2$ (only $\aK$ depends on this and, as mentioned above, $\aK$ is essentially unconstrained by observational constraints on linearly perturbed cosmologies).
However, the $c_{sst}$ bound is highly constraining, since in an expanding universe it demands,
\begin{align} \label{posprior}
& \varphi\varphi \;\; {\rm prior}: &\frac{\bar{G}_{4,X}^2}{\bar{G}_4}  &\leq - \bar G_{4,XX} &\Rightarrow& &\aB &\leq \frac{2 \aT}{1+\aT},
\end{align}
where we have used \eqref{alpha3} as well as the fact that $M^2 = 2 G_4/(1+\aT)$. Note that this expression for $M^2$ holds for the specific model under consideration here, not in general.

\paragraph{$\varphi\chi$ scattering prior:}
The bound from $2 \to 2$ matter-scalar scattering additionally imposes
\begin{align} \label{lumprior}
&\varphi\chi \;\; {\rm prior}:
 &\bar G_{4,X} &\geq 0 &\Rightarrow&
&\aT &\geq 0  \, ,
\end{align}
where $\chi$ is an arbitrary matter field that may live in an entirely separate sector, as discussed above. Phrased in a different way, this bound imposes that the speed of gravitational waves in the EFT \eqref{quarticA} is equal to or larger than the speed of light.
Importantly, note that this bound is therefore orthogonal to some of the subluminality priors that have been considered previously -- see e.g. the seminal work of \cite{Salvatelli:2016mgy} and the more recent \cite{Melville:2019wyy}.
Again we emphasise that both positivity bounds used here were derived on a Lorentz-invariant Minkowski background ($\phi = 0$) -- hence the bars on the left of \eqref{posprior} and \eqref{lumprior} -- while we ultimately rephrase these as bounds on the $\alpha_i$, i.e. as bounds on functions of the cosmological background. As discussed above, we therefore assume that one can port constraints from one background to the other, invoking the covariant nature of \eqref{quarticA} in the process.

\paragraph{Cosmological parameter constraints:}

%
\begin{figure}[t]
\begin{center}
\includegraphics[width=.7\linewidth]{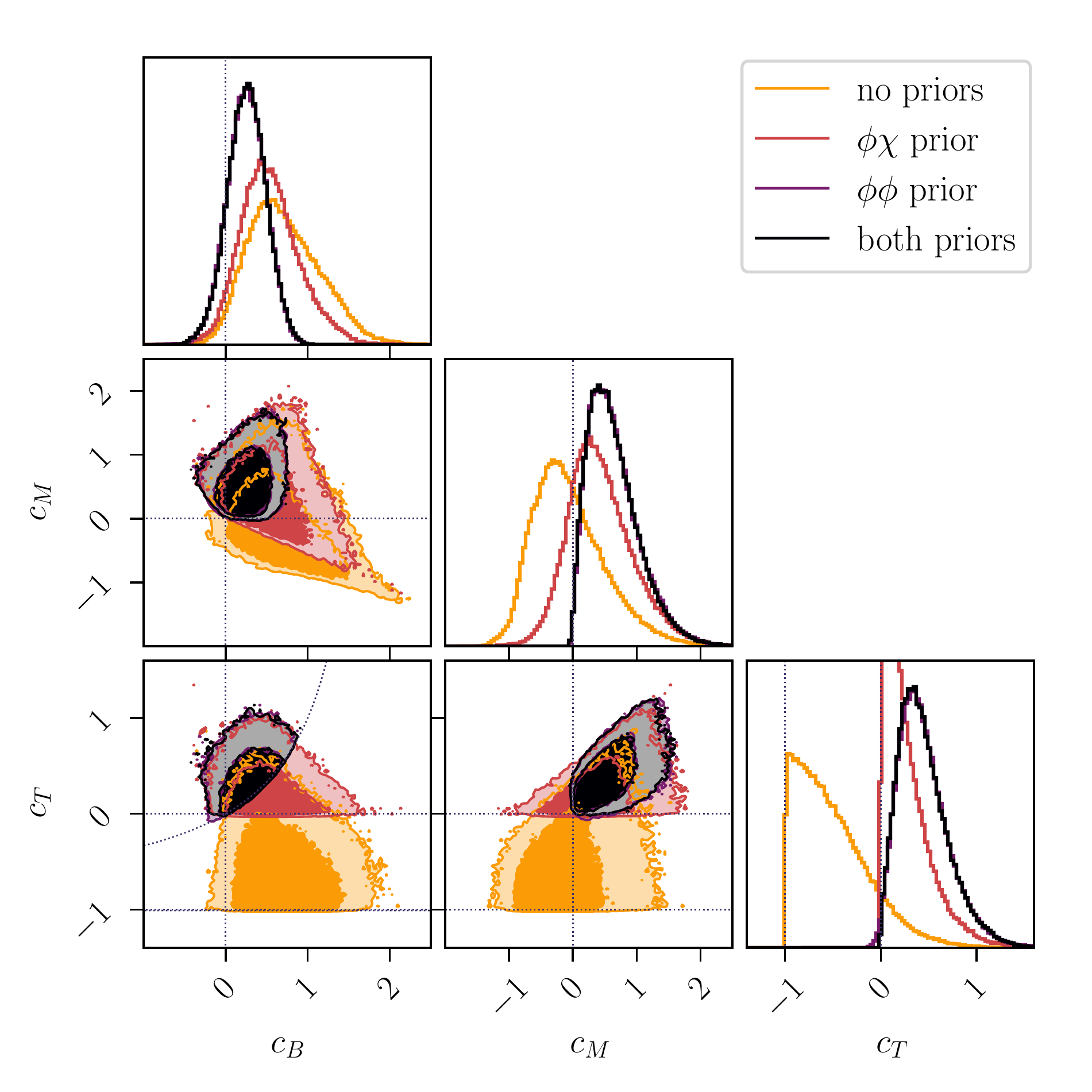}
\end{center}
\caption{Cosmological parameter constraints for the EFT of dark energy considered in \eqref{quarticA}, using different combinations of positivity priors -- see \eqref{posprior} and \eqref{lumprior}. The priors are derived from $\varphi\varphi \to \varphi\varphi$ and $\varphi\chi \to \varphi\chi$ scattering, where $\chi$ denotes matter fields and $\varphi$ the fluctuation of the dark energy field.
Contours mark $68\%$ and $95\%$ confidence intervals, computed using CMB, RSD, BAO and matter power spectrum measurements, and we use $\aI = c_i \Omega_{\rm DE}$ \eqref{oParam}.
Dotted lines mark $c_i=0$ (the GR value), $c_T \geq -1$ (real GW speed) and $c_B < 2 c_T/(1+ c_T)$ (late-time bound from $\varphi\varphi$ scattering). Note that the $\varphi\varphi$ prior constraint is only very marginally enhanced by adding the $\varphi\chi$ prior, see e.g. the $c_T - c_B$ plot for negative $c_T$. Note that the speed of gravitational waves is given by $c^2_{\rm GW}=1+c_T \Omega_{\rm DE}$.
\label{fig1}}
\end{figure}

We can now apply the theoretical priors \eqref{posprior} and \eqref{lumprior}  to cosmological parameter constraints. Doing so in the presence of free functions, such as $G_4(X)$ in \eqref{quarticA}, requires choosing a parametrisation for the freedom encoded in these functions. Instead of choosing a particular ansatz for these functions in the Lagrangian (e.g. a truncated expansion in powers of $X$), we here follow the approach employed by several current Einstein-Boltzmann solvers (see e.g. \cite{Hu:2013twa,Zumalacarregui:2016pph}) and parameterise at the level of the $\aI$. Numerous such parameterisations exist -- see \cite{Noller:2018wyv} and references therein for how these affect constraints and for discussions of their relative merits -- but here we will pick arguably the one most frequently used \cite{Bellini:2014fua} for illustration
\be \label{oParam}
\aI = c_i \Omega_{\rm DE}.
\ee
This parameterises each $\aI$ in terms of just one constant parameter, $c_i$, and is known to accurately capture the evolution of a wide sub-class of Horndeski theories \cite{Pujolas:2011he,Barreira:2014jha}.
Note that, when imposing priors, we will require them to be satisfied at all times, i.e. dynamically throughout the evolution until today {\it as well as} at late times, when $\Omega_{\rm DE} \to 1$ on our $\Lambda$CDM-like background. In the context of \eqref{oParam}, this late time limit yields the strongest bounds on the $c_i$, given the above priors on the $\alpha_i$.

We now compute constraints on the $\aI$ by performing a Markov chain Monte Carlo (MCMC) analysis, using Planck 2015 CMB temperature, CMB lensing and low-$\ell$ polarisation data \cite{Planck-Collaboration:2016af, Planck-Collaboration:2016aa, Planck-Collaboration:2016ae}, baryon acoustic oscillation (BAO) measurements from SDSS/BOSS \cite{Anderson:2014, Ross:2015}, constraints from the SDSS DR4 LRG matter power spectrum shape \cite{Tegmark:2006} and redshift space distortion (RSD) constraints from BOSS and 6dF \cite{Beutler:2012, Samushia:2014}. Note that the constraints presented here are strongly driven by Planck and RSD data, whereas the other data sets do not add significant extra constraining power in our context -- see \cite{Noller:2018wyv} for details regarding the implementation of the relevant likelihoods and related theoretical and observational modelling details.
Using the parametrisation \eqref{oParam}, we can therefore compute constraints on the modified gravity/dark energy parameters $c_{B}, c_M$ and $c_{T}$, marginalising over the standard $\Lambda{\rm CDM}$ parameters $\Omega_{\rm cdm}, \Omega_{\rm b}, \theta_s,A_s,n_s$ and $\tau_{\rm reio}$. The results are shown in Fig.~\ref{fig1} and in Table~\ref{tab1}. 

Focusing on the results as presented in  Fig.~\ref{fig1}, we recover the result of \cite{Melville:2019wyy} that applying the $\varphi\varphi \to \varphi\varphi$ positivity prior significantly reduces the $2\sigma$ volume in parameter space. It effectively rules out $c_B > 1$ and $c_M < 0$, i.e. the running of the Planck mass is always zero or positive here and the amount of `braiding' present is limited. Key for our discussion is that the $\varphi\varphi \to \varphi\varphi$ positivity prior also already introduces a very strong preference for $c_T > 0$ (gravitational waves propagating faster than light). A very small region of parameter space with $c_T < 0$ and $c_B < 0$ is still a good fit to the data, but is dwarfed in volume by the other regions within the $2\sigma$ volume. As such, the added inclusion of the  $\varphi\chi \to \varphi\chi$ positivity prior from scalar-matter scattering only has a very small effect on the combined constraints. Since this prior effectively rules out $c_T < 0$, it only modifies the previous results by ruling out the aforementioned very small (previously remaining) region of parameter space with negative $c_T$.
We present results for the marginalised 1D distributions of the $c_i$ parameters in Table~\ref{tab1}, where we accordingly observe nearly identical such posteriors when both priors are applied and when only the $\varphi\varphi \to \varphi\varphi$ prior is applied,  with only a very small noticeable difference between the two cases for the $c_T$ distribution.
Crucially, however, the scalar-matter prior makes it clear that (assuming a standard UV completion) applying a different subluminality prior on the tensor modes here is not consistent with positivity bounds at large.
Taken together the two positivity priors are therefore remarkably consistent here, with the $\varphi\varphi \to \varphi\varphi$ clearly the stronger bound in the present context.

It is instructive to quantify the relative strength of the (different subsets of) positivity-induced bounds more precisely. We will do so by using $\Delta \equiv \Delta c_B \Delta c_M \Delta c_T$ as a rough measure of the `volume' in parameter space allowed by a given set of constraints, where $\Delta c_i$ denotes the 95\% confidence interval for the posterior of that $c_i$ -- for example, with no priors we have $\Delta c_B = 0.90 + 0.71 = 1.61$. This simple measure is of course not unique, but it will be perfectly sufficient to roughly quantify and compare the relative strength of different combinations of constraints. From table \ref{tab1} we can then see that the viable parameter space volume $\Delta$ shrinks by close to $70\%$ when comparing constraints without any positivity-induced priors vs. constraints where these priors are applied. In addition, there is a noticeable shift towards larger (positive) $c_M$, smaller $c_B$ and larger (positive) $c_T$. That reduction in parameter space volume is near-identical with the one achieved when only applying the $\varphi\varphi \to \varphi\varphi$ positivity prior and is to be compared with a roughly $50\%$ reduction in parameter space volume $\Delta$, when only applying the $\varphi\chi \to \varphi\chi$ positivity prior.
Importantly, the prior from matter-scalar scattering therefore has a significant effect by itself, e.g. preferring larger $c_M$ values than without positivity priors (in addition to the more obvious effect on $c_T$). So, in its own right, it remains a powerful prior to use for computing cosmological data constraints. We therefore caution against ignoring this second prior, because it effectively being pre-empted by the $\varphi\varphi \to \varphi\varphi$ prior here may be a consequence of the specific example model we have chosen. This is especially important, since we expect the scalar-matter prior to generically be linked to $\aT$ (see the related discussion in the appendix), while the nature of the scalar-scalar prior is likely to change more model-dependently. For a related discussion see \cite{Melville:2019wyy}.

Finally, we should point out a number of ways in which the present analysis can be strengthened and extended going forward. First, note that there can be an interesting interplay between the choice of parametrisation for the $\aI$ and the constraining power of theoretical priors in the present context \cite{Kennedy:2020ehn}. Increasingly physically well-motivated and theoretically constrained parameterisations for the $\aI$ should remedy (some of) this modelling uncertainty in the future. In addition, the present analysis can be improved by incorporating further observations and constraints. Examples include adding weak lensing constraints to the analysis along the lines described in \cite{SpurioMancini:2019rxy} or incorporating additional constraints from solar system scales, e.g. recent bounds on $\aM$ using lunar laser ranging \cite{Burrage:2020jkj}.

\begin{figure}
\begin{center}
\includegraphics[width=.7\linewidth]{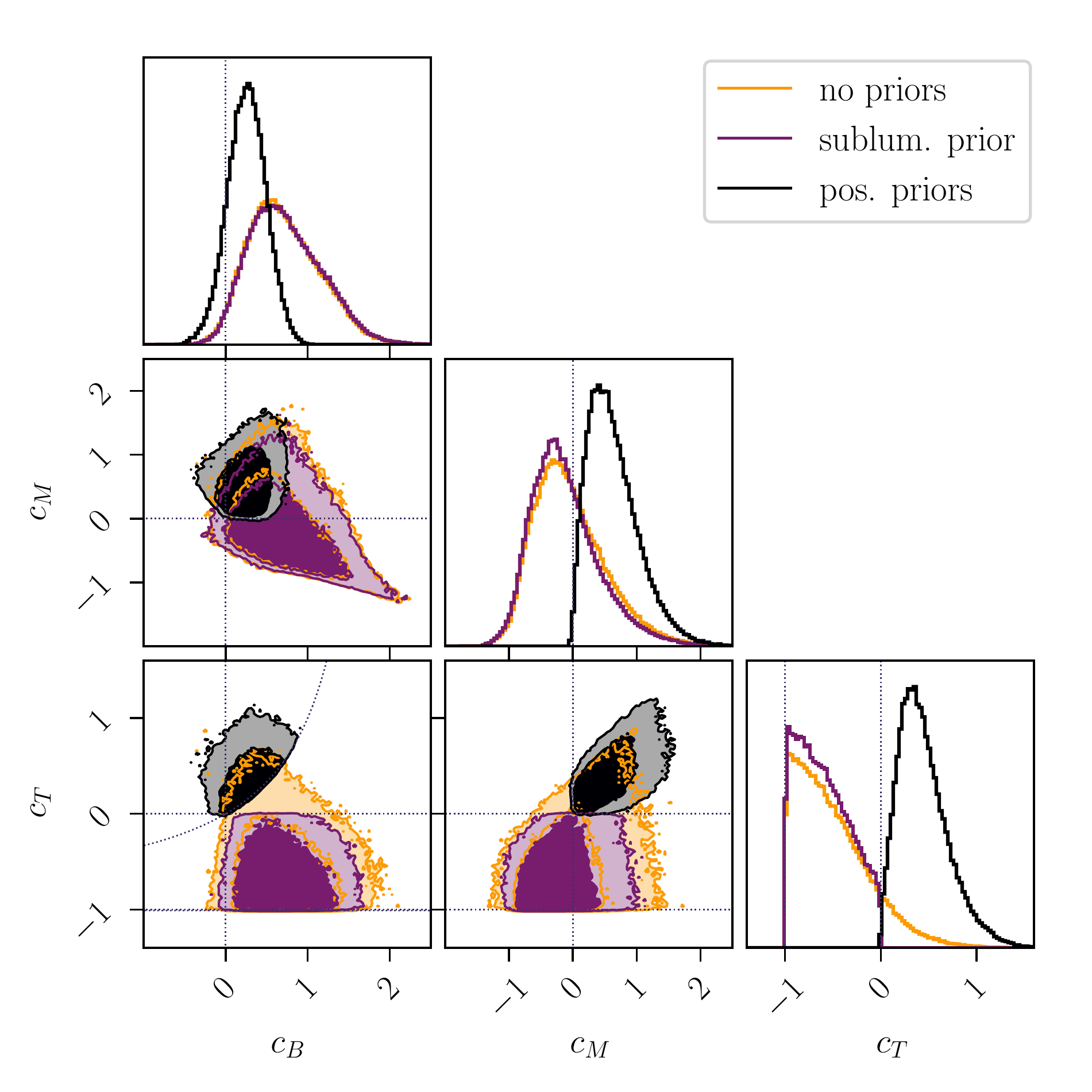}
\end{center}
\caption{Comparison of constraints on the cosmological parameters for the EFT \eqref{quarticA}, where $\alpha_i=c_i \Omega_{\rm DE}$ and we are using either priors inferred from positivity bounds (related to causality, locality and unitarity) or a subluminality prior (which is {\it not} related to causality here). Note that, as we are comparing other priors here, 
gradient and ghost stability priors are applied uniformly for all cases (including 'no priors').
We emphasise that the subluminality prior is incompatible with the positivity bounds shown here. So requiring subluminality (with or without explicitly requiring further stability bounds) leads to drastically different conclusions compared with requiring positivity bounds to be satisfied which include in  themselves causality requirements.
Note that the speed of gravitational waves is given by $c^2_{\rm GW}=1+c_T \Omega_{\rm DE}$.
 \label{fig:Sub_vs_Pos}}
\end{figure}

\paragraph{Comparison with priors relying solely on stability and subluminality:}
%
To highlight the strength of positivity priors when putting constraints on cosmological parameters, it is instructive to compare with what one would have inferred,  had we not used any knowledge from the implementation of our low-energy EFT within its high-energy completion, and had we instead based our priors solely on the stability of the low-energy EFT and subluminality of the dark energy scalar field as well as subluminality of tensor modes.
\\

\noindent $\bullet$ {\bf Stability} of the low-energy EFT requires absence of ghosts and gradients instabilities for both scalar and tensor modes -- see \cite{Kobayashi:2011nu} for general expressions for these requirements in Horndeski scalar-tensor theories.
In terms of the $\alpha_i$, these conditions amounts to \cite{Bellini:2014fua,Zumalacarregui:2016pph}
\begin{align}
&\text{scalar modes}:    &\text{no ghost}:& \; {\cal D} \equiv \alpha_K + \tfrac{3}{2}\alpha_B^2 > 0,  &\text{grad. stab.}:& \; c_s^2 \geq 0, \nonumber \\
&\text{tensor modes}:    &\text{no ghost}:& \; M^2 > 0,   &\text{grad. stab.}:& \; \alpha_T \geq -1\,, \label{ghost_conditions}
\end{align}
where
\begin{align} \label{cs2}
{\cal D}c_s^2 &= (2-\aB)\left(\tfrac{1}{2}\aB(1+\aT) + \aM - \aT - \frac{\dot H}{H^2}\right) + \frac{\dot{\alpha}_B}{H} - \frac{3(\rho_{\rm tot} + p_{\rm tot})}{H^2M^2}.
\end{align}
Here $\rho_{\rm tot}$ and $p_{\rm tot}$ are the total energy density and pressure in the universe. Ghost and gradient stability priors are uniformly applied for all the constraints we are showing throughout this paper\footnote{As an aside, note that these priors are effectively imposed automatically for the setup we are considering, which importantly includes the ansatz \eqref{oParam}.  This is because the data here exclude almost all regions, where such instabilities are present, by themselves, regardless of whether ghost- and gradient-stability priors are explicitly imposed or not \cite{Noller:2018eht,Noller:2018wyv}.}. 
\smi{We note in passing that, while the positivity bound on $G_{4,X}$ requires gravitational waves to travel faster than matter in this model, the scalar fluctuations may travel faster or slower than matter.}

Note that the above ghost and gradient stability criteria are derived by considering scalar and tensor modes propagating on an FLRW background.
One may of course complement this by specifying other backgrounds/vacua the EFT in question should also describe. Here we are implicitly requiring this to be the case for the flat Minkowski vacuum, given that we are porting positivity bounds from there.  Another example of interest is demanding the stability of propagating scalar modes on backgrounds sourced by massive binary systems \cite{Creminelli:2019kjy} -- the resulting theoretical priors can then have a strong effect on cosmological parameter constraints as well \cite{Noller:2020afd}. We leave a further exploration of these issues for future work, but stress that the interplay between more exhaustive (future) sets of stability and positivity priors is likely to allow us to extract increasingly tight constraints.
\\

{\noindent $\bullet$ {\bf Subluminality vs Causality.} It is worth emphasizing from the outset that the notion of subluminality in this gravitational EFT is not directly linked with that of causality and may sometimes be orthogonal to it. Indeed the notion of causality is intrinsically linked to the speed of propagation of information which is related to the front velocity, i.e. the high (or even infinite) frequency limit of the phase velocity \cite{Milonni,Brillouin,deRham:2014zqa}. Implementing conditions related to causality therefore requires some knowledge of the high-energy behaviour (by definition beyond the regime of validity of the EFT) which is precisely what is encoded in the positivity bounds. Within a low-energy EFT, the notion of causality is much more subtle to identify \cite{deRham:2019ctd,deRham:2020zyh,Reall:2021voz}.
In the EFT \eqref{quarticA}, from \eqref{cs2}, subluminality of the scalar mode
can always be achieved by requiring a sufficiently large kineticity $\aK$ (and hence ${\cal D}$). Since (as discussed above) $\aK$ is effectively unconstrained by observations here, this means a subluminality prior for the speed of scalar perturbations does not lead to any significant constraints in our context\footnote{It does place one rather weak constraint. Namely, a lower bound on $\aK$, which from \eqref{cs2} itself depends on the other $\aI$.}.
Requiring (sub)luminality of the tensor modes, on the other hand, would impose $\alpha_T \leq 0$. Taken together with the gradient stability prior for tensor modes, at late times when $\Omega_{\rm DE} \to 1$, this means the prior boundaries would  effectively become $-1 \leq c_T \leq 0$.

Given the exact same cosmological data as used previously, a (sub)luminality prior
would therefore lead to rather different conclusions than the priors from positivity we discussed above. This is illustrated in Fig.~\ref{fig:Sub_vs_Pos}, where constraints using our positivity priors (that include the requirement of causality) are compared with the constraints one would have inferred had we instead focused on (sub)luminality priors (which in this case do no go hand in hand with causality). We again emphasise that ghost and gradient stability priors are uniformly implemented for all cases, so differences between the cases shown are solely due to the other priors as discussed above.
This illustrates the importance of carefully choosing priors based on theoretical consistency when constraining EFT parameters from data.

\section{Discussion}\label{sec:discussion}

In this paper, we have considered the effects of incorporating a novel positivity bound into the analysis of cosmological scalar-tensor theories. Specifically, this results from including bounds derived from interactions with any other matter fields (which are known to interact at the very least gravitationally).
In order to illustrate the effect of additional positivity priors derived in the presence of any matter fields, we extended the analysis of \cite{Melville:2019wyy}, considering a specific (quartic and shift-symmetric) dark energy EFT and computing the corresponding positivity priors derived from scalar-scalar and scalar-matter scattering. While the qualitatively new prior from matter-scalar scattering does significantly tighten constraints in comparison with those derived using observational data only, we also find that for the specific theory considered here, this new prior is effectively pre-empted by the known one from scalar-scalar scattering. Phrased differently, this means the two priors are highly consistent with one another here. We emphasise, however, that both priors in principle constrain the parameter space in completely different ways -- the matter-scalar bound requires that the speed of tensor modes  is strictly greater than the speed of any matter field (including photons) -- so should both be taken into account when going beyond the example theory used here in the future.

With an eye on such future surveys, it will be very interesting to see whether additional positivity bounds on dark energy (continue to) display a high degree of consistency, both for the example model considered here and for more generic EFTs. At the moment, the high level of consistency between  the two sets of positivity bounds appears to be suggestive that once priors from a causal and unitary high energy completion are accounted for, data naturally folds itself in a region of parameter space that then remains highly consistent with other such considerations.
However, it is possible that in the future, additional positivity bounds could  restrict the parameter space in highly orthogonal ways, which would then provide even more powerful constraints, effectively  ruling out large classes of dark energy candidates.
\\

\acknowledgments

\vspace{-0.1in}
\noindent  We would like to thank Andrew J.~Tolley for initial collaboration and useful insights.
JN is supported by an STFC Ernest Rutherford Fellowship, grant reference ST/S004572/1.
SM is supported by an UKRI Stephen Hawking Fellowship (EP/T017481/1) and partially by STFC consolidated grants ST/P000681/1 and ST/T000694/1.
CdR acknowledges financial support provided by the European Union's Horizon 2020 Research Council grant 724659 MassiveCosmo ERC-2016-COG, by STFC grants ST/P000762/1 and ST/T000791/1, by the Royal Society through a Wolfson Research Merit Award, by the Simons Foundation award ID 555326 under the Simons Foundation's Origins of the Universe initiative, `\textit{Cosmology Beyond Einstein's Theory}' and by the Simons Investigator award 690508.
In deriving the results of this paper, we have used: CLASS \cite{Blas:2011rf},  corner \cite{corner}, hi\_class \cite{Zumalacarregui:2016pph,Bellini:2019syt}, MontePyton \cite{Audren:2012wb,Brinckmann:2018cvx} and xAct \cite{xAct}.

\appendix
\section{Scattering amplitudes}
\label{app:A}

In this Appendix we review the amplitudes which lead to the constraints \eqref{posbounds} and \eqref{posbounds2}
comparing the positivity requirement with relative speed of tensors to matter fields, $c_{\rm GW}^2 / c_m^2$.
We start by expanding the metric $g_{\mu\nu} = \bar{g}_{\mu\nu} + h_{\mu\nu}/\mpl$ and the scalar field $\phi = \bar{\phi} + \varphi$ in \eqref{quarticA}, and use an overbar $\G_n$ to denote functions evaluated on the flat background. $\chi$ is used throughout to denote matter fields.

 \subsection{Scattering on a flat background}

We first consider the amplitudes about a flat background and without loss of generality, we set $\bar{G}_{2, X} = 1/2$ to canonically normalise the $\varphi$ propagator. The most important point to notice is that for the theory considered in \eqref{quarticA}, the coupling between $h$ and $\varphi$ to leading order in the vertex $h \varphi \varphi$ has the precise same structure as the  graviton kinetic's term, so that whenever the emission of two $\varphi$'s is mediated by $h_{\mu\nu}$ there is no pole in the amplitude (up to the $\mathcal{O} ( 1/\mpl )$ at which we are working). There is therefore no $t$-channel pole to affect the applicability of our bounds to this order.

\subsubsection*{$\bullet \ \varphi \varphi \to \varphi \varphi$ scattering}
The leading contributions to the $\varphi \varphi \to \varphi\varphi$ amplitude are given by
\FloatBarrier
\begin{figure}[h!t]
\centering
		\begin{tikzpicture}[baseline=-0.6cm]
			\begin{feynman}
				\vertex (a1);
				\vertex [below=0.5cm of a1] (b1);
				\vertex [below=0.5cm of b1] (c1);
				\vertex [left=0.5cm of a1] (a2);
				\vertex [below=0.5cm of a2] (b2);
				\vertex [above left=0.4cm of b2] (b2lt);
				\vertex [below left=0.4cm of b2] (b2lb);
				\vertex [above right=0.4cm of b2] (b2rt);
				\vertex [below right=0.4cm of b2] (b2rb);
				\vertex [below=0.5cm of b2] (c2);
				\vertex [left=0.5cm of a2] (a3);
				\vertex [below=0.5cm of a3] (b3);
				\vertex [below=0.5cm of b3] (c3);
				\vertex [left=0.5cm of a3] (a4);
				\vertex [below=0.5cm of a4] (b4);
				\vertex [below=0.5cm of b4] (c4);
 				\node [below=0.1cm of a2, blob];
				
				\diagram*{
				(a3) -- [scalar] (b2lt),
				(c3) -- [scalar] (b2lb),
				(a1) -- [scalar] (b2rt),
				(c1) -- [scalar] (b2rb),
				};				
			\end{feynman}		
		\end{tikzpicture}
		\qquad = \quad
		\begin{tikzpicture}[baseline=-0.6cm]
			\begin{feynman}
				\vertex (a1);
				\vertex [below=0.5cm of a1] (b1);
				\vertex [below=0.5cm of b1] (c1);
				\vertex [left=0.5cm of a1] (a2);
				\vertex [below=0.5cm of a2] (b2);
				\vertex [below=0.5cm of b2] (c2);
				\vertex [left=0.5cm of a2] (a3);
				\vertex [below=0.5cm of a3] (b3);
				\vertex [below=0.5cm of b3] (c3);
				\vertex [left=0.5cm of a3] (a4);
				\vertex [below=0.5cm of a4] (b4);
				\vertex [below=0.5cm of b4] (c4);
				\node at (-0.5,  -1.5) {$ \frac{ \G_{4,XX} }{\Lambda_3^{6}} $};
				
				\diagram*{
				(a3) -- [scalar] (b2),
				(c3) -- [scalar] (b2),
				(a1) -- [scalar] (b2),
				(c1) -- [scalar] (b2),
				};				
			\end{feynman}		
		\end{tikzpicture}
		\qquad + \qquad
 		\begin{tikzpicture}[baseline=-0.6cm]
			\begin{feynman}
				\vertex (a1);
				\vertex [below=0.5cm of a1] (b1);
				\vertex [below=0.5cm of b1] (c1);
				\vertex [left=0.5cm of a1] (a2);
				\vertex [below=0.5cm of a2] (b2);
				\vertex [left=0.15cm of b2] (b2l);
				\vertex [above right=0.2cm of b2] (b2rt);
				\vertex [below right=0.2cm of b2] (b2rb);
				\vertex [below=0.5cm of b2] (c2);
				\vertex [left=0.8cm of a2] (a3);
				\vertex [below=0.5cm of a3] (b3);
				\vertex [right=0.15cm of b3] (b3r);
				\vertex [above left=0.2cm of b3] (b3lt);
				\vertex [below left=0.2cm of b3] (b3lb);
				\vertex [below=0.5cm of b3] (c3);
				\vertex [left=0.5cm of a3] (a4);
				\vertex [below=0.5cm of a4] (b4);
				\vertex [below=0.5cm of b4] (c4);
 				\node at (-0.8,  -1.5) {$\frac{ \G_{4,X}^2 }{\Lambda_3^{6}}$};
				
				\diagram*{
				(a4) -- [scalar] (b3),
				(c4) -- [scalar] (b3),
				(b3) -- [photon] (b2),
				(b2) -- [scalar] (a1),
				(b2) -- [scalar] (c1),
				};				
			\end{feynman}		
		\end{tikzpicture}
		\qquad + \qquad  perm.  \\
		\qquad\qquad\; + \quad
		\begin{tikzpicture}[baseline=-0.6cm]
			\begin{feynman}
				\vertex (a1);
				\vertex [below=0.5cm of a1] (b1);
				\vertex [below=0.5cm of b1] (c1);
				\vertex [left=0.5cm of a1] (a2);
				\vertex [below=0.5cm of a2] (b2);
				\node   [below=0.425cm of a2, dot] (d);
				\vertex [below=0.5cm of b2] (c2);
				\vertex [left=0.5cm of a2] (a3);
				\vertex [below=0.5cm of a3] (b3);
				\vertex [below=0.5cm of b3] (c3);
				\vertex [left=0.5cm of a3] (a4);
				\vertex [below=0.5cm of a4] (b4);
				\vertex [below=0.5cm of b4] (c4);
				\node at (-0.5,  -1.5) {$\frac{ \G_{2, XX} }{\mpl \Lambda_3^{3}}$};
				
				\diagram*{
				(a3) -- [scalar] (b2),
				(c3) -- [scalar] (b2),
				(a1) -- [scalar] (b2),
				(c1) -- [scalar] (b2),
				};				
			\end{feynman}		
		\end{tikzpicture}
		\qquad + \qquad
 		\begin{tikzpicture}[baseline=-0.6cm]
			\begin{feynman}
				\vertex (a1);
				\vertex [below=0.5cm of a1] (b1);
				\vertex [below=0.5cm of b1] (c1);
				\vertex [left=0.5cm of a1] (a2);
				\vertex [below=0.5cm of a2] (b2);
				\node   [below=0.425cm of a2, dot] (d);
				\vertex [left=0.15cm of b2] (b2l);
				\vertex [above right=0.2cm of b2] (b2rt);
				\vertex [below right=0.2cm of b2] (b2rb);
				\vertex [below=0.5cm of b2] (c2);
				\vertex [left=0.8cm of a2] (a3);
				\vertex [below=0.5cm of a3] (b3);
				\vertex [right=0.15cm of b3] (b3r);
				\vertex [above left=0.2cm of b3] (b3lt);
				\vertex [below left=0.2cm of b3] (b3lb);
				\vertex [below=0.5cm of b3] (c3);
				\vertex [left=0.5cm of a3] (a4);
				\vertex [below=0.5cm of a4] (b4);
				\vertex [below=0.5cm of b4] (c4);
 				\node at (-0.8,  -1.5) {$\frac{ \G_{4,X} }{2\mpl \Lambda_3^{3}}$};
				
				\diagram*{
				(a4) -- [scalar] (b3),
				(c4) -- [scalar] (b3),
				(b3) -- [photon] (b2),
				(b2) -- [scalar] (a1),
				(b2) -- [scalar] (c1),
				};				
			\end{feynman}		
		\end{tikzpicture}
		\qquad + \qquad perm.
\end{figure}
\FloatBarrier
\noindent where dashed lines indicate the scalar field $\varphi$ while wiggly lines indicate the graviton propagator.
To leading order, the resulting amplitude to the $\varphi \varphi \to \varphi \varphi$ process is therefore,
\begin{align}
\mathcal{A}^{\varphi \varphi \to \varphi \varphi}  &= 4!  \frac{ \bar{G}_{4, XX}   +  \G_{4,X}^2 /   \G_4 }{\Lambda_3^6}  \; \frac{stu}{4}
+  \frac{ \G_{2,XX} }{\Lambda_2^4}  \left( s^2 + t^2 + u^2  \right)  -  \frac{ \G_{4,X}}{ \G_4 \Lambda_2^4 }  \left(  su + st + u t  \right)  \, ,
\end{align}
\noindent which at large $s$ takes the form
\begin{equation}
\mathcal{A} (s,t) =  c_{ss} ~ \frac{s^2}{\Lambda_2^4}  + c_{sst} ~ \frac{ s^2 t}{\Lambda_3^6} +  ... \; ,
\label{Ast}
\end{equation}
with,
\begin{align}
  c_{sst} = - 6 \left(  \bar{G}_{4,XX} + \bar{G}_{4,X}^2  /  \G_4 \right)  \; , \;\;\;\; c_{ss} = 2 \bar{G}_{2, XX}  +   \bar{G}_{4,X}    /  \G_4\,.
\end{align}
This gives the two positivity bounds studied in \cite{Melville:2019wyy},
\begin{align}
\bar G_{2,XX} &\geq  - \bar{G}_{4,X} \frac{ \G_{2,X} }{\G_4}  \,  , &  \bar G_{4,XX} + \frac{\bar{G}_{4,X}^2}{\G_4}  &\leq 0  \, .
\end{align}
\subsubsection*{$\bullet \ \varphi h \to \varphi h$ scattering}
For the $\varphi h \to \varphi h$ amplitude, the only non-zero diagrams are,
\begin{figure}[h!t]
\centering
		\begin{tikzpicture}[baseline=-0.6cm]
			\begin{feynman}
				\vertex (a1);
				\vertex [below=0.5cm of a1] (b1);
				\vertex [below=0.5cm of b1] (c1);
				\vertex [left=0.5cm of a1] (a2);
				\vertex [below=0.5cm of a2] (b2);
				\vertex [above left=0.4cm of b2] (b2lt);
				\vertex [below left=0.4cm of b2] (b2lb);
				\vertex [above right=0.4cm of b2] (b2rt);
				\vertex [below right=0.4cm of b2] (b2rb);
				\vertex [below=0.5cm of b2] (c2);
				\vertex [left=0.5cm of a2] (a3);
				\vertex [below=0.5cm of a3] (b3);
				\vertex [below=0.5cm of b3] (c3);
				\vertex [left=0.5cm of a3] (a4);
				\vertex [below=0.5cm of a4] (b4);
				\vertex [below=0.5cm of b4] (c4);
 				\node [below=0.1cm of a2, blob];
				
				\diagram*{
				(a3) -- [scalar] (b2lt),
				(c3) -- [photon] (b2lb),
				(a1) -- [scalar] (b2rt),
				(c1) -- [photon] (b2rb),
				};				
			\end{feynman}		
		\end{tikzpicture}
		\qquad = \quad
		\begin{tikzpicture}[baseline=-0.6cm]
			\begin{feynman}
				\vertex (a1);
				\vertex [below=0.5cm of a1] (b1);
				\vertex [below=0.5cm of b1] (c1);
				\vertex [left=0.5cm of a1] (a2);
				\vertex [below=0.5cm of a2] (b2);
				\node   [below=0.425cm of a2, dot] (d);
				\vertex [below=0.5cm of b2] (c2);
				\vertex [left=0.5cm of a2] (a3);
				\vertex [below=0.5cm of a3] (b3);
				\vertex [below=0.5cm of b3] (c3);
				\vertex [left=0.5cm of a3] (a4);
				\vertex [below=0.5cm of a4] (b4);
				\vertex [below=0.5cm of b4] (c4);
				\node at (-0.5,  -1.5) {$\frac{ \G_{4,X} }{\mpl \Lambda_3^{3}}$};
				
				\diagram*{
				(a3) -- [scalar] (b2),
				(c3) -- [photon] (b2),
				(a1) -- [scalar] (b2),
				(c1) -- [photon] (b2),
				};				
			\end{feynman}		
		\end{tikzpicture}
		\qquad + \qquad
 		\begin{tikzpicture}[baseline=-0.6cm]
			\begin{feynman}
				\vertex (a1);
				\vertex [below=0.5cm of b1] (c1);
				\vertex [left=0.5cm of a1] (a2);
				\vertex [below=0.5cm of a2] (b2);
				\vertex [below=0.5cm of b2] (c2);
				\vertex [left=0.5cm of a2] (a3);
				\vertex [below=0.5cm of a3] (b3);
				\vertex [below=0.5cm of b3] (c3);
				\vertex [left=0.5cm of a3] (a4);
				\vertex [below=0.5cm of a4] (b4);
				\vertex [below=0.5cm of b4] (c4);
				\vertex [below=0.25cm of a2] (a2l);
				\vertex [above=0.25cm of c2] (c2u);
 				\node at (-0.5,  -1.5) {$\frac{ \G_{4,X} \G_{4} }{\mpl \Lambda_3^{3}}$};
 				\node   [above=0.15cm of c2, dot] (d);
				
				\diagram*{
				(a3) -- [scalar] (a2l) -- [scalar] (a1),
				(c3) -- [photon] (c2u) -- [photon] (c1),
				(c2u) -- [photon] (a2l),
				};				
			\end{feynman}		
		\end{tikzpicture}
		\qquad + \qquad perm.\,,
\end{figure}
\FloatBarrier
\noindent which exactly cancel, so $\mathcal{A}_{\varphi h \to \varphi h} = 0$ at this order.
\subsubsection*{$\bullet \ \varphi \chi \to \varphi \chi$ scattering}
Including matter fields, $\chi$, now gives an additional amplitude for $\varphi \chi \to \varphi \chi$. In the frame in which \eqref{quarticA} is defined, there is no direct coupling between any of the matter fields and the dark energy field $\varphi$, but a scattering process is mediated by gravitational exchange,
\FloatBarrier
\begin{figure}[h]
\centering
		\begin{tikzpicture}[baseline=-0.6cm]
			\begin{feynman}
				\vertex (a1);
				\vertex [below=0.5cm of a1] (b1);
				\vertex [below=0.5cm of b1] (c1);
				\vertex [left=0.5cm of a1] (a2);
				\vertex [below=0.5cm of a2] (b2);
				\vertex [left=0.15cm of b2] (b2l);
				\vertex [above right=0.2cm of b2] (b2rt);
				\vertex [below right=0.2cm of b2] (b2rb);
				\vertex [below=0.5cm of b2] (c2);
				\vertex [left=0.8cm of a2] (a3);
				\vertex [below=0.5cm of a3] (b3);
				\vertex [right=0.15cm of b3] (b3r);
				\vertex [above left=0.2cm of b3] (b3lt);
				\vertex [below left=0.2cm of b3] (b3lb);
				\vertex [below=0.5cm of b3] (c3);
				\vertex [left=0.5cm of a3] (a4);
				\vertex [below=0.5cm of a4] (b4);
				\vertex [below=0.5cm of b4] (c4);
 				\node at (0.2,  0.1) {$\varphi$};
  				\node at (0.2,  -1.1) {$\varphi$};
    				\node at (-2.0,  -0.5) {$T^{\mu\nu}$};
    				\node at (-0.8,  -0.8) {\footnotesize $h_{\mu\nu}$};
  				\node [left=0.0cm of b3, crossed dot];
				
				\diagram*{
				(b3) -- [photon] (b2),
				(b2) -- [scalar, momentum = \footnotesize $p_\mu$] (a1),
				(b2) -- [scalar] (c1),
				};				
			\end{feynman}		
		\end{tikzpicture}
		\qquad $=$ \qquad  $ \frac{ 2 \bar{G}_{4, X}  }{ \G_4 \mpl \Lambda_3^3} T^{\mu\nu} p_\mu p_\nu $
\end{figure}
\FloatBarrier
\noindent Again we see the graviton propagator cancelling against the $h$ derivatives in the vertex factor, leaving only an effectively contact coupling between the current $T^{\mu\nu}$ of $\chi$ and the momentum $p_\mu$ of one of the $\varphi$'s (by momentum conservation, the momentum of the other $\varphi$ can be written as $-p_\mu - p^{(T)}_\mu$, but $p^{(T)}_\mu T^{\mu\nu} = 0$ since $T^{\mu\nu}$ is conserved). For example for a light scalar matter field (with canonical kinetic term and neglecting self-interactions),
\begin{align}
 T^{\mu\nu} &= \nabla^\mu \chi \nabla^\nu \chi - \frac{1}{2} g^{\mu\nu} \nabla^\alpha \chi \nabla_\alpha \chi \;\; ,
 \label{eqn:Tscalar}
\end{align}
this gives a single ($t$-channel) diagram,
\FloatBarrier
\begin{figure}[h]
\centering
		\begin{tikzpicture}[baseline=-0.6cm]
			\begin{feynman}
				\vertex (a1);
				\vertex [below=0.5cm of a1] (b1);
				\vertex [below=0.5cm of b1] (c1);
				\vertex [left=0.5cm of a1] (a2);
				\vertex [below=0.5cm of a2] (b2);
				\vertex [left=0.15cm of b2] (b2l);
				\vertex [above right=0.2cm of b2] (b2rt);
				\vertex [below right=0.2cm of b2] (b2rb);
				\vertex [below=0.5cm of b2] (c2);
				\vertex [left=0.8cm of a2] (a3);
				\vertex [below=0.5cm of a3] (b3);
				\vertex [right=0.15cm of b3] (b3r);
				\vertex [above left=0.2cm of b3] (b3lt);
				\vertex [below left=0.2cm of b3] (b3lb);
				\vertex [below=0.5cm of b3] (c3);
				\vertex [left=0.5cm of a3] (a4);
				\vertex [below=0.5cm of a4] (b4);
				\vertex [below=0.5cm of b4] (c4);
 				\node at (0.6,  0.1) {$\varphi (p_2)$};
  				\node at (0.6,  -1.1) {$\varphi (p_4)$};
 				\node at (-2.4,  0.1) {$\chi (p_1)$};
 				\node at (-2.4,  -1.1) {$\chi (p_3)$};
    				\node at (-0.9,  -0.8) {\footnotesize $h_{\mu\nu}$};
				
				\diagram*{
				(a4) -- [] (b3),
				(c4) -- [] (b3),
				(b3) -- [photon] (b2),
				(b2) -- [scalar] (a1),
				(b2) -- [scalar] (c1),
				};				
			\end{feynman}		
		\end{tikzpicture}
		\qquad $=$ \qquad  $ \frac{ \bar{G}_{4, X}  }{2 \Lambda_2^4} ( s^2  + u^2 - t^2 )$
\end{figure}
\FloatBarrier
\noindent and so the amplitude $\mathcal{A}_{\varphi \chi \to \varphi \chi}$ takes the form \eqref{Ast} with,
\begin{align}
 c_{ss} =   \bar{G}_{4,X}   \;  .
 \label{eqn:Ascalar}
\end{align}
This gives a new positivity bound on the function $G_4$ (expanded about a flat background),
\begin{align}
 \bar{G}_{4,X}  \geq 0 \, ,
 \label{eqn:new_pos_bound}
\end{align}
which is qualitatively different to the previous bounds found from $\varphi \varphi$ (and $\varphi h$) elastic scattering without matter fields.

Note that one obtains the same bound for vector matter fields, $\chi_\mu$, since,
\begin{align}
 T^{\mu\nu} &= F^{\mu \alpha} F^{\nu}_{\;\; \alpha} - \frac{1}{4} g^{\mu\nu} F^{\alpha\beta} F_{\alpha\beta} \;\; \,,
\end{align}
where $F_{\mu\nu} = \partial_{[\mu} \chi_{\nu]}$ (we assume that $\chi$ is Abelian and minimally coupled to gravity) gives forward-limit helicity amplitudes,
\begin{align}
 \mathcal{A}^{\varphi \chi \to \varphi \chi}_{0 \pm \to 0 \pm} = + \frac{ 4 \G_{4,X}  }{ \G_4 \Lambda_2^4} \; s^2 .
\end{align}
When both a scalar and vector (ingredients of the Standard Model) couple minimally to the metric indicated in  \eqref{quarticA} then positivity requires \eqref{eqn:new_pos_bound}\footnote{
Coupling higher spin fields to gravity is difficult (e.g. when the spin is $\geq 5/2$, the gauge symmetry required to decouple longitudinal modes is deformed non-trivially by the curvature \cite{Aragone:1979hx, Porrati:1993in, Cucchieri:1994tx}), and for half-integer spin requires a spin connection---but the speeds of spin-0 and spin-1 fields are constrained to be $\leq c_T$. What matters for us is that, as soon as there is a single spin-0 or spin-1 matter field in the universe coupled to gravity like $+h_{\mu\nu} T^{\mu\nu}/2/\mpl$, then $\bar{G}_{4,X}$ must be positive.
}.

\subsubsection*{Sound speeds on a cosmological background}

On a non-trivial background, the leading contributions to the second derivatives of the metric perturbations in de Donder gauge and focusing on the tensor modes, the relevant contributions are of the form
\begin{eqnarray}
\L_{hh}^{2}\sim  \bar G_4 h^\mn \bar \Box h_\mn+\frac{\Lambda_2^4 \bar G_{4,X} }{\Lambda_3^6\mpl^2}\partial^\alpha \bar \phi \partial^\beta \bar \phi h^\mn \partial_\alpha \partial_\beta h_\mn+\cdots\,.
\end{eqnarray}
Diagonalizing the metric at a point, then for a time-like background where $\partial \bar \phi \sim \delta^0_\mu$ so that $\bar X>0$, the speed of the tensor modes is given by
\begin{eqnarray}
\frac{c_{\rm GW}^2}{c_m^2}=\frac{\bar G_4}{\bar G_4-2 \bar X \bar G_{4,X}}\,,
\end{eqnarray}
where we have included the speed $c_m$ of matter fields minimally coupled to the metric $g_\mn$. In this frame, $c_m=1$.
We conclude that the speed of gravitational waves is larger than the speed $c_m$ of matter fields minimally coupled to the metric $g_\mn$ if $\bar G_{4,X}$ is positive.
Had we instead consider a space-like background where $\partial \bar \phi \sim \delta^0_i$ so that $\bar X<0$ (assuming again we diagonalized the metric at that point),  then the speed of the tensor modes is given by
\begin{eqnarray}
\frac{c_{\rm GW}^2}{c_m^2}=\frac{\bar G_4-2 \bar X \bar G_{4,X}}{\bar G_4}\,,
\end{eqnarray}
which is again superluminal if $\bar G_{4,X}$ is positive.

More generally, had we included  the quintic Horndeski term,
\begin{align}
 \mathcal{L}_5&=   \mpl^2 G_{5} G_{\mu\nu}\Phi^{\mu\nu} -\tfrac{1}{6}\Lambda_2^4  G_{5,X}  ([\Phi]^3 -3[\Phi][\Phi^2]+2[\Phi^3])  ,
\end{align}
the tensor sound speed on a cosmological background would then have been,
\begin{align}
 c_{\rm GW}^2  = 1 + \frac{ 2 \bar {X} }{ M^2 } \left( 2\bar{G}_{4,X} - \bar{G}_{5,\phi} + ( H \partial_t \bar{\phi} - \partial_t^2 \bar{\phi} ) \bar{G}_{5,X}     \right) \, ,
\end{align}
where now $M^2 = 2 \bar{G}_4 + \bar{X} \left( -4  \bar{G}_{4,X} + \G_{5,\phi} - 2 H \partial_t \bar{\phi} \, \bar{G}_{5,X}  \right) $ is the effective Planck mass (assumed positive). On the other hand, $\varphi \chi$ scattering on a flat background is not sensitive to $H$, with an amplitude given by,
\begin{align}
c_{ss} = \G_{4,X} - \tfrac{1}{2} \G_{5, \phi}
\end{align}
and so positivity again requires $c_{\rm GW}^2 \geq c_m^2$, where $c_m$ is the speed of matter fields minimally coupled to the metric $g_\mn$, up to $\mathcal{O} (H)$ corrections for Horndeski-type theories.

\subsection{Einstein frame}

It is instructive to show how the same positivity constraints arise in the Einstein frame. This is both a useful consistency check (in particular it removes the need to ever discussing about the graviton pole), and it also makes it clear why it is the $\varphi \chi$ amplitude which is responsible for the previous sound speed relation (since in the Einstein frame $c_{\rm GW}^2 =1$ the effect of the $\bar \phi$ background is to lower $c_m$, which is precisely the statement encoded in the $\varphi \chi$ amplitude).

The equations of motion for $\phi$ and for $h_{\mu\nu}$ from $S[h_{\mu\nu}, \phi] + S_m [h_{\mu\nu}, \chi]$ can be written as,
\begin{align}
2 \G_{2,X}  \phi_\mu^\mu - 4 \frac{ g_4 }{\Lambda_3^6} \,  \phi_\mu^{[\mu} \phi_\alpha^\alpha \phi_\rho^{\rho]}   &= + \frac{2 \bar{G}_{4,X} }{ \G_4 \Lambda_2^4 } T^\mu_{\;\; \nu} \phi_{\mu}^\nu    \label{eqn:Horn1} \\
 \delta_{\nu \beta \sigma}^{\mu \alpha \rho} \nabla_{\alpha} \nabla^\beta \left[  h_{\rho}^{\;\; \sigma} - \frac{2 \bar{G}_{4,X} }{ \G_4 \Lambda_3^3} \phi_\rho  \phi^{\sigma}   \right]  &=  - \frac{1}{\G_4 \mpl} T^{\mu}_{\nu}  \label{eqn:Horn2}
\end{align}
where $g_4 =  \bar{G}_{4,XX}  +  \bar{G}_{4,X}^2/\bar{G_4} $.
This shows that the mixing between $\phi$ and $h_{\mu \nu}$ can be removed at this order by redefining the metric fluctuation,
\begin{equation}
h_{\rho  \sigma} =  h^{(E)}_{\rho \sigma} + \frac{2 \bar{G}_{4,X} }{ \G_4 \Lambda_3^3}  \varphi_\rho  \varphi_\sigma \, ,
 \label{eqn:fieldredef}
\end{equation}
and thus $2 \bar{G}_{4,X}/ \G_4$ plays the role of a disformal coupling in the Einstein frame.
In particular, from the canonical $h_{\mu\nu} T^{\mu\nu}$ coupling in the Jordan frame, the matter is now coupled to,
\begin{equation}
 \frac{1}{2 \mpl} h_{\mu\nu} T^{\mu\nu} =  \frac{1}{2 \mpl} h^{(E)}_{\mu\nu} T^{\mu\nu} + \frac{\G_{4,X} }{ \G_4 \Lambda_2^4} \varphi_\mu \varphi_\nu T^{\mu\nu} \, \, ,
\end{equation}
in the Einstein frame.

\subsubsection*{Scattering on a flat background}

In this frame, $h^{(E)}_{\mu\nu}$ does not couple directly to $\varphi$  at this order, but there is instead a contact interaction $T^{\mu\nu} \nabla_{\mu} \varphi \nabla_{\nu} \varphi$ which is responsible for the $\varphi \chi$ scattering,
 \FloatBarrier
\begin{figure}[h]
\centering
		\begin{tikzpicture}[baseline=-0.6cm]
			\begin{feynman}
				\vertex (a1);
				\vertex [below=0.5cm of a1] (b1);
				\vertex [below=0.5cm of b1] (c1);
				\vertex [left=0.5cm of a1] (a2);
				\vertex [below=0.5cm of a2] (b2);
				\vertex [left=0.15cm of b2] (b2l);
				\vertex [above right=0.2cm of b2] (b2rt);
				\vertex [below right=0.2cm of b2] (b2rb);
				\vertex [below=0.5cm of b2] (c2);
				\vertex [left=0.5cm of a2] (a3);
				\vertex [below=0.5cm of a3] (b3);
				\vertex [right=0.15cm of b3] (b3r);
				\vertex [above left=0.2cm of b3] (b3lt);
				\vertex [below left=0.2cm of b3] (b3lb);
				\vertex [below=0.5cm of b3] (c3);
				\vertex [left=0.5cm of a3] (a4);
				\vertex [below=0.5cm of a4] (b4);
				\vertex [below=0.5cm of b4] (c4);
 				\node at (0.2,  0.1) {$\varphi$};
  				\node at (0.2,  -1.1) {$\varphi$};
 				\node at (-1.2,  -0.4) {$T^{\mu\nu}$};
 				\node [left=0.0cm of b2, crossed dot];
				
				\diagram*{
				(b2) -- [scalar, momentum=\footnotesize $p_\mu$] (a1),
				(b2) -- [scalar] (c1),
				};				
			\end{feynman}		
		\end{tikzpicture}		\qquad $=$ \qquad  $ \frac{ 2 \G_{4, X}  }{ \G_4 \Lambda_2^4} T^{\mu\nu} p_\mu p_\nu $\,.
\end{figure}
\FloatBarrier
\noindent For instance, for the scalar matter field \eqref{eqn:Tscalar}, there is again a single diagram,
\begin{figure}[h]
\centering
		\begin{tikzpicture}[baseline=-0.6cm]
			\begin{feynman}
				\vertex (a1);
				\vertex [below=0.5cm of a1] (b1);
				\vertex [below=0.5cm of b1] (c1);
				\vertex [left=0.5cm of a1] (a2);
				\vertex [below=0.5cm of a2] (b2);
				\vertex [left=0.15cm of b2] (b2l);
				\vertex [above right=0.2cm of b2] (b2rt);
				\vertex [below right=0.2cm of b2] (b2rb);
				\vertex [below=0.5cm of b2] (c2);
				\vertex [left=0.5cm of a2] (a3);
				\vertex [below=0.5cm of a3] (b3);
				\vertex [right=0.15cm of b3] (b3r);
				\vertex [above left=0.2cm of b3] (b3lt);
				\vertex [below left=0.2cm of b3] (b3lb);
				\vertex [below=0.5cm of b3] (c3);
				\vertex [left=0.5cm of a3] (a4);
				\vertex [below=0.5cm of a4] (b4);
				\vertex [below=0.5cm of b4] (c4);
 				\node at (0.6,  0.1) {$\varphi (p_2)$};
  				\node at (0.6,  -1.1) {$\varphi (p_4)$};
 				\node at (-1.6,  0.1) {$\chi (p_1)$};
 				\node at (-1.6,  -1.1) {$\chi (p_3)$};
				
				\diagram*{
				(a3) -- [] (b2),
				(c3) -- [] (b2),
				(b2) -- [scalar] (a1),
				(b2) -- [scalar] (c1),
				};				
			\end{feynman}		
		\end{tikzpicture}
		\qquad $=$ \qquad  $ \frac{ G_{4, X}  }{2 \Lambda_2^4} ( s^2  + u^2 - t^2 )$\,,
\end{figure}
\FloatBarrier
\noindent which reproduces the amplitude \eqref{eqn:Ascalar} and positivity bound \eqref{eqn:new_pos_bound} (as obviously expected, since a field redefinition like \eqref{eqn:fieldredef} leaves $S$-matrix elements unchanged).

\subsubsection*{Sound speeds on a cosmological background}

In the Einstein frame, the equation of motion for $h^{(E)}_{\mu\nu}$ is unaffected by the background of $\bar{\phi}$, and tensor modes therefore propagate luminally just as in GR, $c_{\rm GW}=1$. However, matter now couples to an effective metric, which on this background reads,
\begin{align}
 g_{\mu\nu} = \bar{g}^{(E)}_{\mu\nu}  + \frac{2 \bar{G}_{4,X} }{ \G_4 \Lambda_2^4 }  \partial_\mu \bar{\phi} \partial_\nu \bar{\phi} \; ,
\end{align}
and consequently the sound speed of all matter fields is now shifted by $\bar{\phi}$ relative to $c_{\rm GW}$, by an amount
\begin{align}
\label{eq:cc1}
 \frac{c_m^2}{c_{\rm GW}^2} = 1  -  \frac{2 \bar{X} \bar{G}_{4,X} }{ \G_4 }  \; .
\end{align}
where $\bar{X} =  \left( \partial_t \bar{\phi} \right)^2 /  \Lambda_2^4$ on a time-like background.

In fact, for any constant disformal coupling to matter in the Einstein frame,
\begin{align}
 g_{\mu\nu} = g^{(E)}_{\mu\nu}  +  \frac{  \partial_\mu \bar \phi \partial_\nu \bar\phi  }{ \mathcal{M}^4 }
\end{align}
positivity demands that $\mathcal{M}^4 \geq 0$, and therefore the speed of matter relative to the speed of $\delta g^{(E)}_{\mu\nu}$ fluctuations,
\begin{align}
\frac{c_m^2}{c_{\rm GW}^2} = 1  - \frac{  \bar{X} }{ \mathcal{M}^4 }
\end{align}
is constrained to be less than one on time-like backgrounds for $\bar\phi$ (for which $\bar{X} \geq 0$). Similarly as in the previous argument, when dealing with a space-like background, the relation \eqref{eq:cc1} is inverted and we still recover the same outcome $c_m^2\le c_{\rm GW}^2$.

\bibliographystyle{JHEP}
\bibliography{pos_matter_couplings}

\providecommand{\href}[2]{#2}\begingroup\raggedright\begin{thebibliography}{100}

\bibitem{Gubitosi:2012hu}
G.~Gubitosi, F.~Piazza and F.~Vernizzi, \emph{{The Effective Field Theory of
  Dark Energy}},
  \href{http://dx.doi.org/10.1088/1475-7516/2013/02/032}{\emph{JCAP} {\bfseries
  1302} (2013) 032}, [\href{https://arxiv.org/abs/1210.0201}{{\ttfamily
  1210.0201}}].

\bibitem{Bloomfield:2012ff}
J.~K. Bloomfield, E.~E. Flanagan, M.~Park and S.~Watson, \emph{{Dark energy or
  modified gravity? An effective field theory approach}},
  \href{http://dx.doi.org/10.1088/1475-7516/2013/08/010}{\emph{JCAP} {\bfseries
  1308} (2013) 010}, [\href{https://arxiv.org/abs/1211.7054}{{\ttfamily
  1211.7054}}].

\bibitem{Gleyzes:2014rba}
J.~Gleyzes, D.~Langlois and F.~Vernizzi, \emph{{A unifying description of dark
  energy}}, \href{http://dx.doi.org/10.1142/S021827181443010X}{\emph{Int. J.
  Mod. Phys.} {\bfseries D23} (2015) 1443010},
  [\href{https://arxiv.org/abs/1411.3712}{{\ttfamily 1411.3712}}].

\bibitem{Bellini:2014fua}
E.~Bellini and I.~Sawicki, \emph{{Maximal freedom at minimum cost: linear
  large-scale structure in general modifications of gravity}},
  \href{http://dx.doi.org/10.1088/1475-7516/2014/07/050}{\emph{JCAP} {\bfseries
  1407} (2014) 050}, [\href{https://arxiv.org/abs/1404.3713}{{\ttfamily
  1404.3713}}].

\bibitem{Gleyzes:2013ooa}
J.~Gleyzes, D.~Langlois, F.~Piazza and F.~Vernizzi, \emph{{Essential Building
  Blocks of Dark Energy}},
  \href{http://dx.doi.org/10.1088/1475-7516/2013/08/025}{\emph{JCAP} {\bfseries
  08} (2013) 025}, [\href{https://arxiv.org/abs/1304.4840}{{\ttfamily
  1304.4840}}].

\bibitem{Kase:2014cwa}
R.~Kase and S.~Tsujikawa, \emph{{Effective field theory approach to modified
  gravity including Horndeski theory and Horava-Lifshitz gravity}},
  \href{http://dx.doi.org/10.1142/S0218271814430081}{\emph{Int. J. Mod. Phys.}
  {\bfseries D23} (2015) 1443008},
  [\href{https://arxiv.org/abs/1409.1984}{{\ttfamily 1409.1984}}].

\bibitem{DeFelice:2015isa}
A.~De~Felice, K.~Koyama and S.~Tsujikawa, \emph{{Observational signatures of
  the theories beyond Horndeski}},
  \href{http://dx.doi.org/10.1088/1475-7516/2015/05/058}{\emph{JCAP} {\bfseries
  1505} (2015) 058}, [\href{https://arxiv.org/abs/1503.06539}{{\ttfamily
  1503.06539}}].

\bibitem{Langlois:2017mxy}
D.~Langlois, M.~Mancarella, K.~Noui and F.~Vernizzi, \emph{{Effective
  Description of Higher-Order Scalar-Tensor Theories}},
  \href{http://dx.doi.org/10.1088/1475-7516/2017/05/033}{\emph{JCAP} {\bfseries
  1705} (2017) 033}, [\href{https://arxiv.org/abs/1703.03797}{{\ttfamily
  1703.03797}}].

\bibitem{Frusciante:2019xia}
N.~Frusciante and L.~Perenon, \emph{{Effective field theory of dark energy: A
  review}}, \href{http://dx.doi.org/10.1016/j.physrep.2020.02.004}{\emph{Phys.
  Rept.} {\bfseries 857} (2020) 1--63},
  [\href{https://arxiv.org/abs/1907.03150}{{\ttfamily 1907.03150}}].

\bibitem{Renevey:2020tvr}
C.~Renevey, J.~Kennedy and L.~Lombriser, \emph{{Parameterised post-Newtonian
  formalism for the effective field theory of dark energy via screened
  reconstructed Horndeski theories}},
  \href{http://dx.doi.org/10.1088/1475-7516/2020/12/032}{\emph{JCAP} {\bfseries
  12} (2020) 032}, [\href{https://arxiv.org/abs/2006.09910}{{\ttfamily
  2006.09910}}].

\bibitem{Lagos:2016wyv}
M.~Lagos, T.~Baker, P.~G. Ferreira and J.~Noller, \emph{{A general theory of
  linear cosmological perturbations: scalar-tensor and vector-tensor
  theories}},
  \href{http://dx.doi.org/10.1088/1475-7516/2016/08/007}{\emph{JCAP} {\bfseries
  1608} (2016) 007}, [\href{https://arxiv.org/abs/1604.01396}{{\ttfamily
  1604.01396}}].

\bibitem{Lagos:2017hdr}
M.~Lagos, E.~Bellini, J.~Noller, P.~G. Ferreira and T.~Baker, \emph{{A general
  theory of linear cosmological perturbations: stability conditions, the
  quasistatic limit and dynamics}},
  \href{http://dx.doi.org/10.1088/1475-7516/2018/03/021}{\emph{JCAP} {\bfseries
  1803} (2018) 021}, [\href{https://arxiv.org/abs/1711.09893}{{\ttfamily
  1711.09893}}].

\bibitem{Noller:2018wyv}
J.~Noller and A.~Nicola, \emph{{Cosmological parameter constraints for
  Horndeski scalar-tensor gravity}},
  \href{http://dx.doi.org/10.1103/PhysRevD.99.103502}{\emph{Phys. Rev. D}
  {\bfseries 99} (2019) 103502},
  [\href{https://arxiv.org/abs/1811.12928}{{\ttfamily 1811.12928}}].

\bibitem{BelliniParam}
E.~{Bellini}, A.~J. {Cuesta}, R.~{Jimenez} and L.~{Verde}, \emph{{Constraints
  on deviations from {$\Lambda$}CDM within Horndeski gravity}},
  \href{http://dx.doi.org/10.1088/1475-7516/2016/02/053}{\emph{JCAP} {\bfseries
  2} (Feb., 2016) 053}, [\href{https://arxiv.org/abs/1509.07816}{{\ttfamily
  1509.07816}}].

\bibitem{Hu:2013twa}
B.~Hu, M.~Raveri, N.~Frusciante and A.~Silvestri, \emph{{Effective Field Theory
  of Cosmic Acceleration: an implementation in CAMB}},
  \href{http://dx.doi.org/10.1103/PhysRevD.89.103530}{\emph{Phys. Rev. D}
  {\bfseries 89} (2014) 103530},
  [\href{https://arxiv.org/abs/1312.5742}{{\ttfamily 1312.5742}}].

\bibitem{Raveri:2014cka}
M.~Raveri, B.~Hu, N.~Frusciante and A.~Silvestri, \emph{{Effective Field Theory
  of Cosmic Acceleration: constraining dark energy with CMB data}},
  \href{http://dx.doi.org/10.1103/PhysRevD.90.043513}{\emph{Phys. Rev.}
  {\bfseries D90} (2014) 043513},
  [\href{https://arxiv.org/abs/1405.1022}{{\ttfamily 1405.1022}}].

\bibitem{Gleyzes:2015rua}
J.~Gleyzes, D.~Langlois, M.~Mancarella and F.~Vernizzi, \emph{{Effective Theory
  of Dark Energy at Redshift Survey Scales}},
  \href{http://dx.doi.org/10.1088/1475-7516/2016/02/056}{\emph{JCAP} {\bfseries
  1602} (2016) 056}, [\href{https://arxiv.org/abs/1509.02191}{{\ttfamily
  1509.02191}}].

\bibitem{Kreisch:2017uet}
C.~D. Kreisch and E.~Komatsu, \emph{{Cosmological Constraints on Horndeski
  Gravity in Light of GW170817}},
  \href{https://arxiv.org/abs/1712.02710}{{\ttfamily 1712.02710}}.

\bibitem{Zumalacarregui:2016pph}
M.~Zumalac\'arregui, E.~Bellini, I.~Sawicki, J.~Lesgourgues and P.~G. Ferreira,
  \emph{{hi_class: Horndeski in the Cosmic Linear Anisotropy Solving System}},
  \href{http://dx.doi.org/10.1088/1475-7516/2017/08/019}{\emph{JCAP} {\bfseries
  1708} (2017) 019}, [\href{https://arxiv.org/abs/1605.06102}{{\ttfamily
  1605.06102}}].

\bibitem{Alonso:2016suf}
D.~Alonso, E.~Bellini, P.~G. Ferreira and M.~Zumalac\'arregui,
  \emph{{Observational future of cosmological scalar-tensor theories}},
  \href{http://dx.doi.org/10.1103/PhysRevD.95.063502}{\emph{Phys. Rev.}
  {\bfseries D95} (2017) 063502},
  [\href{https://arxiv.org/abs/1610.09290}{{\ttfamily 1610.09290}}].

\bibitem{Arai:2017hxj}
S.~Arai and A.~Nishizawa, \emph{{Generalized framework for testing gravity with
  gravitational-wave propagation. II. Constraints on Horndeski theory}},
  \href{http://dx.doi.org/10.1103/PhysRevD.97.104038}{\emph{Phys. Rev.}
  {\bfseries D97} (2018) 104038},
  [\href{https://arxiv.org/abs/1711.03776}{{\ttfamily 1711.03776}}].

\bibitem{Frusciante:2018jzw}
N.~Frusciante, S.~Peirone, S.~Casas and N.~A. Lima, \emph{{The road ahead of
  Horndeski: cosmology of surviving scalar-tensor theories}},
  \href{https://arxiv.org/abs/1810.10521}{{\ttfamily 1810.10521}}.

\bibitem{Reischke:2018ooh}
R.~Reischke, A.~S. Mancini, B.~M. Schafer and P.~M. Merkel,
  \emph{{Investigating scalar-tensor-gravity with statistics of the cosmic
  large-scale structure}},  \href{https://arxiv.org/abs/1804.02441}{{\ttfamily
  1804.02441}}.

\bibitem{Mancini:2018qtb}
A.~Spurio~Mancini, R.~Reischke, V.~Pettorino, B.~M. Schafer and
  M.~Zumalac\'arregui, \emph{{Testing (modified) gravity with 3D and
  tomographic cosmic shear}},
  \href{http://dx.doi.org/10.1093/mnras/sty2092}{\emph{Mon. Not. Roy. Astron.
  Soc.} {\bfseries 480} (2018) 3725},
  [\href{https://arxiv.org/abs/1801.04251}{{\ttfamily 1801.04251}}].

\bibitem{Brando:2019xbv}
G.~Brando, F.~T. Falciano, E.~V. Linder and H.~E.~S. Velten, \emph{{Modified
  gravity away from a $\Lambda$CDM background}},
  \href{http://dx.doi.org/10.1088/1475-7516/2019/11/018}{\emph{JCAP} {\bfseries
  11} (2019) 018}, [\href{https://arxiv.org/abs/1904.12903}{{\ttfamily
  1904.12903}}].

\bibitem{Arjona:2019rfn}
R.~Arjona, W.~Cardona and S.~Nesseris, \emph{{Designing Horndeski and the
  effective fluid approach}},
  \href{http://dx.doi.org/10.1103/PhysRevD.100.063526}{\emph{Phys. Rev. D}
  {\bfseries 100} (2019) 063526},
  [\href{https://arxiv.org/abs/1904.06294}{{\ttfamily 1904.06294}}].

\bibitem{Raveri:2019mxg}
M.~Raveri, \emph{{Reconstructing Gravity on Cosmological Scales}},
  \href{http://dx.doi.org/10.1103/PhysRevD.101.083524}{\emph{Phys. Rev. D}
  {\bfseries 101} (2020) 083524},
  [\href{https://arxiv.org/abs/1902.01366}{{\ttfamily 1902.01366}}].

\bibitem{Perenon:2019dpc}
L.~Perenon, J.~Bel, R.~Maartens and A.~de~la Cruz-Dombriz, \emph{{Optimising
  growth of structure constraints on modified gravity}},
  \href{http://dx.doi.org/10.1088/1475-7516/2019/06/020}{\emph{JCAP} {\bfseries
  06} (2019) 020}, [\href{https://arxiv.org/abs/1901.11063}{{\ttfamily
  1901.11063}}].

\bibitem{SpurioMancini:2019rxy}
A.~Spurio~Mancini, F.~K\"ohlinger, B.~Joachimi, V.~Pettorino, B.~M. Sch\"afer,
  R.~Reischke et~al., \emph{{KiDS + GAMA: constraints on horndeski gravity from
  combined large-scale structure probes}},
  \href{http://dx.doi.org/10.1093/mnras/stz2581}{\emph{Mon. Not. Roy. Astron.
  Soc.} {\bfseries 490} (2019) 2155--2177},
  [\href{https://arxiv.org/abs/1901.03686}{{\ttfamily 1901.03686}}].

\bibitem{Bonilla:2019mbm}
A.~Bonilla, R.~D'Agostino, R.~C. Nunes and J.~C.~N. de~Araujo, \emph{{Forecasts
  on the speed of gravitational waves at high $z$}},
  \href{http://dx.doi.org/10.1088/1475-7516/2020/03/015}{\emph{JCAP} {\bfseries
  03} (2020) 015}, [\href{https://arxiv.org/abs/1910.05631}{{\ttfamily
  1910.05631}}].

\bibitem{Baker:2020apq}
T.~Baker and I.~Harrison, \emph{{Constraining Scalar-Tensor Modified Gravity
  with Gravitational Waves and Large Scale Structure Surveys}},
  \href{http://dx.doi.org/10.1088/1475-7516/2021/01/068}{\emph{JCAP} {\bfseries
  01} (2021) 068}, [\href{https://arxiv.org/abs/2007.13791}{{\ttfamily
  2007.13791}}].

\bibitem{Joudaki:2020shz}
S.~Joudaki, P.~G. Ferreira, N.~A. Lima and H.~A. Winther, \emph{{Testing
  Gravity on Cosmic Scales: A Case Study of Jordan-Brans-Dicke Theory}},
  \href{https://arxiv.org/abs/2010.15278}{{\ttfamily 2010.15278}}.

\bibitem{Noller:2020lav}
J.~Noller, L.~Santoni, E.~Trincherini and L.~G. Trombetta, \emph{{Scalar-tensor
  cosmologies without screening}},
  \href{http://dx.doi.org/10.1088/1475-7516/2021/01/045}{\emph{JCAP} {\bfseries
  01} (2021) 045}, [\href{https://arxiv.org/abs/2008.08649}{{\ttfamily
  2008.08649}}].

\bibitem{Noller:2020afd}
J.~Noller, \emph{{Cosmological constraints on dark energy in light of
  gravitational wave bounds}},
  \href{http://dx.doi.org/10.1103/PhysRevD.101.063524}{\emph{Phys. Rev. D}
  {\bfseries 101} (2020) 063524},
  [\href{https://arxiv.org/abs/2001.05469}{{\ttfamily 2001.05469}}].

\bibitem{Adams:2006sv}
A.~Adams, N.~Arkani-Hamed, S.~Dubovsky, A.~Nicolis and R.~Rattazzi,
  \emph{{Causality, analyticity and an IR obstruction to UV completion}},
  \href{http://dx.doi.org/10.1088/1126-6708/2006/10/014}{\emph{JHEP} {\bfseries
  10} (2006) 014}, [\href{https://arxiv.org/abs/hep-th/0602178}{{\ttfamily
  hep-th/0602178}}].

\bibitem{Jenkins:2006ia}
A.~Jenkins and D.~O'Connell, \emph{{The Story of O: Positivity constraints in
  effective field theories}},
  \href{https://arxiv.org/abs/hep-th/0609159}{{\ttfamily hep-th/0609159}}.

\bibitem{Adams:2008hp}
A.~Adams, A.~Jenkins and D.~O'Connell, \emph{{Signs of analyticity in fermion
  scattering}},  \href{https://arxiv.org/abs/0802.4081}{{\ttfamily 0802.4081}}.

\bibitem{Nicolis:2009qm}
A.~Nicolis, R.~Rattazzi and E.~Trincherini, \emph{{Energy's and amplitudes'
  positivity}}, \href{http://dx.doi.org/10.1007/JHEP05(2010)095,
  10.1007/JHEP11(2011)128}{\emph{JHEP} {\bfseries 05} (2010) 095},
  [\href{https://arxiv.org/abs/0912.4258}{{\ttfamily 0912.4258}}].

\bibitem{Bellazzini:2014waa}
B.~Bellazzini, L.~Martucci and R.~Torre, \emph{{Symmetries, Sum Rules and
  Constraints on Effective Field Theories}},
  \href{http://dx.doi.org/10.1007/JHEP09(2014)100}{\emph{JHEP} {\bfseries 09}
  (2014) 100}, [\href{https://arxiv.org/abs/1405.2960}{{\ttfamily 1405.2960}}].

\bibitem{Bellazzini:2015cra}
B.~Bellazzini, C.~Cheung and G.~N. Remmen, \emph{{Quantum Gravity Constraints
  from Unitarity and Analyticity}},
  \href{http://dx.doi.org/10.1103/PhysRevD.93.064076}{\emph{Phys. Rev.}
  {\bfseries D93} (2016) 064076},
  [\href{https://arxiv.org/abs/1509.00851}{{\ttfamily 1509.00851}}].

\bibitem{Baumann:2015nta}
D.~Baumann, D.~Green, H.~Lee and R.~A. Porto, \emph{{Signs of Analyticity in
  Single-Field Inflation}},
  \href{http://dx.doi.org/10.1103/PhysRevD.93.023523}{\emph{Phys. Rev.}
  {\bfseries D93} (2016) 023523},
  [\href{https://arxiv.org/abs/1502.07304}{{\ttfamily 1502.07304}}].

\bibitem{Cheung:2016yqr}
C.~Cheung and G.~N. Remmen, \emph{{Positive Signs in Massive Gravity}},
  \href{http://dx.doi.org/10.1007/JHEP04(2016)002}{\emph{JHEP} {\bfseries 04}
  (2016) 002}, [\href{https://arxiv.org/abs/1601.04068}{{\ttfamily
  1601.04068}}].

\bibitem{Bonifacio:2016wcb}
J.~Bonifacio, K.~Hinterbichler and R.~A. Rosen, \emph{{Positivity constraints
  for pseudolinear massive spin-2 and vector Galileons}},
  \href{http://dx.doi.org/10.1103/PhysRevD.94.104001}{\emph{Phys. Rev.}
  {\bfseries D94} (2016) 104001},
  [\href{https://arxiv.org/abs/1607.06084}{{\ttfamily 1607.06084}}].

\bibitem{deRham:2017avq}
C.~de~Rham, S.~Melville, A.~J. Tolley and S.-Y. Zhou, \emph{{Positivity bounds
  for scalar field theories}},
  \href{http://dx.doi.org/10.1103/PhysRevD.96.081702}{\emph{Phys. Rev.}
  {\bfseries D96} (2017) 081702(R)},
  [\href{https://arxiv.org/abs/1702.06134}{{\ttfamily 1702.06134}}].

\bibitem{deRham:2017zjm}
C.~de~Rham, S.~Melville, A.~J. Tolley and S.-Y. Zhou, \emph{{UV complete me:
  Positivity Bounds for Particles with Spin}},
  \href{http://dx.doi.org/10.1007/JHEP03(2018)011}{\emph{JHEP} {\bfseries 03}
  (2018) 011}, [\href{https://arxiv.org/abs/1706.02712}{{\ttfamily
  1706.02712}}].

\bibitem{deRham:2018qqo}
C.~de~Rham, S.~Melville, A.~J. Tolley and S.-Y. Zhou, \emph{{Positivity Bounds
  for Massive Spin-1 and Spin-2 Fields}},
  \href{https://arxiv.org/abs/1804.10624}{{\ttfamily 1804.10624}}.

\bibitem{Bellazzini:2016xrt}
B.~Bellazzini, \emph{{Softness and amplitudes\textquoteright{} positivity for
  spinning particles}},
  \href{http://dx.doi.org/10.1007/JHEP02(2017)034}{\emph{JHEP} {\bfseries 02}
  (2017) 034}, [\href{https://arxiv.org/abs/1605.06111}{{\ttfamily
  1605.06111}}].

\bibitem{deRham:2017imi}
C.~de~Rham, S.~Melville, A.~J. Tolley and S.-Y. Zhou, \emph{{Massive Galileon
  Positivity Bounds}},
  \href{http://dx.doi.org/10.1007/JHEP09(2017)072}{\emph{JHEP} {\bfseries 09}
  (2017) 072}, [\href{https://arxiv.org/abs/1702.08577}{{\ttfamily
  1702.08577}}].

\bibitem{deRham:2017xox}
C.~de~Rham, S.~Melville and A.~J. Tolley, \emph{{Improved Positivity Bounds and
  Massive Gravity}},
  \href{http://dx.doi.org/10.1007/JHEP04(2018)083}{\emph{JHEP} {\bfseries 04}
  (2018) 083}, [\href{https://arxiv.org/abs/1710.09611}{{\ttfamily
  1710.09611}}].

\bibitem{deRham:2019ctd}
C.~de~Rham and A.~J. Tolley, \emph{{Speed of gravity}},
  \href{http://dx.doi.org/10.1103/PhysRevD.101.063518}{\emph{Phys. Rev. D}
  {\bfseries 101} (2020) 063518},
  [\href{https://arxiv.org/abs/1909.00881}{{\ttfamily 1909.00881}}].

\bibitem{deRham:2020zyh}
C.~de~Rham and A.~J. Tolley, \emph{{Causality in curved spacetimes: The speed
  of light and gravity}},
  \href{http://dx.doi.org/10.1103/PhysRevD.102.084048}{\emph{Phys. Rev. D}
  {\bfseries 102} (2020) 084048},
  [\href{https://arxiv.org/abs/2007.01847}{{\ttfamily 2007.01847}}].

\bibitem{Alberte:2020jsk}
L.~Alberte, C.~de~Rham, S.~Jaitly and A.~J. Tolley, \emph{{Positivity Bounds
  and the Massless Spin-2 Pole}},
  \href{https://arxiv.org/abs/2007.12667}{{\ttfamily 2007.12667}}.

\bibitem{Dvali:2010jz}
G.~Dvali, G.~F. Giudice, C.~Gomez and A.~Kehagias, \emph{{UV-Completion by
  Classicalization}},
  \href{http://dx.doi.org/10.1007/JHEP08(2011)108}{\emph{JHEP} {\bfseries 08}
  (2011) 108}, [\href{https://arxiv.org/abs/1010.1415}{{\ttfamily 1010.1415}}].

\bibitem{Dvali:2010ns}
G.~Dvali and D.~Pirtskhalava, \emph{{Dynamics of Unitarization by
  Classicalization}},
  \href{http://dx.doi.org/10.1016/j.physletb.2011.03.054}{\emph{Phys. Lett.}
  {\bfseries B699} (2011) 78--86},
  [\href{https://arxiv.org/abs/1011.0114}{{\ttfamily 1011.0114}}].

\bibitem{Dvali:2011nj}
G.~Dvali, \emph{{Classicalize or not to Classicalize?}},
  \href{https://arxiv.org/abs/1101.2661}{{\ttfamily 1101.2661}}.

\bibitem{Dvali:2011th}
G.~Dvali, C.~Gomez and A.~Kehagias, \emph{{Classicalization of Gravitons and
  Goldstones}}, \href{http://dx.doi.org/10.1007/JHEP11(2011)070}{\emph{JHEP}
  {\bfseries 11} (2011) 070},
  [\href{https://arxiv.org/abs/1103.5963}{{\ttfamily 1103.5963}}].

\bibitem{Vikman:2012bx}
A.~Vikman, \emph{{Suppressing Quantum Fluctuations in Classicalization}},
  \href{http://dx.doi.org/10.1209/0295-5075/101/34001}{\emph{EPL} {\bfseries
  101} (2013) 34001}, [\href{https://arxiv.org/abs/1208.3647}{{\ttfamily
  1208.3647}}].

\bibitem{Kovner:2012yi}
A.~Kovner and M.~Lublinsky, \emph{{Classicalization and Unitarity}},
  \href{http://dx.doi.org/10.1007/JHEP11(2012)030}{\emph{JHEP} {\bfseries 11}
  (2012) 030}, [\href{https://arxiv.org/abs/1207.5037}{{\ttfamily 1207.5037}}].

\bibitem{Keltner:2015xda}
L.~Keltner and A.~J. Tolley, \emph{{UV properties of Galileons: Spectral
  Densities}},  \href{https://arxiv.org/abs/1502.05706}{{\ttfamily
  1502.05706}}.

\bibitem{Horndeski:1974wa}
G.~W. Horndeski, \emph{{Second-order scalar-tensor field equations in a
  four-dimensional space}},
  \href{http://dx.doi.org/10.1007/BF01807638}{\emph{Int. J. Theor. Phys.}
  {\bfseries 10} (1974) 363--384}.

\bibitem{Deffayet:2009wt}
C.~Deffayet, G.~Esposito-Farese and A.~Vikman, \emph{{Covariant Galileon}},
  \href{http://dx.doi.org/10.1103/PhysRevD.79.084003}{\emph{Phys. Rev.}
  {\bfseries D79} (2009) 084003},
  [\href{https://arxiv.org/abs/0901.1314}{{\ttfamily 0901.1314}}].

\bibitem{Pirtskhalava:2015nla}
D.~Pirtskhalava, L.~Santoni, E.~Trincherini and F.~Vernizzi, \emph{{Weakly
  Broken Galileon Symmetry}},
  \href{http://dx.doi.org/10.1088/1475-7516/2015/09/007}{\emph{JCAP} {\bfseries
  1509} (2015) 007}, [\href{https://arxiv.org/abs/1505.00007}{{\ttfamily
  1505.00007}}].

\bibitem{Luty:2003vm}
M.~A. Luty, M.~Porrati and R.~Rattazzi, \emph{{Strong interactions and
  stability in the DGP model}},
  \href{http://dx.doi.org/10.1088/1126-6708/2003/09/029}{\emph{JHEP} {\bfseries
  09} (2003) 029}, [\href{https://arxiv.org/abs/hep-th/0303116}{{\ttfamily
  hep-th/0303116}}].

\bibitem{Nicolis:2004qq}
A.~Nicolis and R.~Rattazzi, \emph{{Classical and quantum consistency of the DGP
  model}}, \href{http://dx.doi.org/10.1088/1126-6708/2004/06/059}{\emph{JHEP}
  {\bfseries 06} (2004) 059},
  [\href{https://arxiv.org/abs/hep-th/0404159}{{\ttfamily hep-th/0404159}}].

\bibitem{deRham:2010eu}
C.~de~Rham and A.~J. Tolley, \emph{{DBI and the Galileon reunited}},
  \href{http://dx.doi.org/10.1088/1475-7516/2010/05/015}{\emph{JCAP} {\bfseries
  05} (2010) 015}, [\href{https://arxiv.org/abs/1003.5917}{{\ttfamily
  1003.5917}}].

\bibitem{Burrage:2010cu}
C.~Burrage, C.~de~Rham, D.~Seery and A.~J. Tolley, \emph{{Galileon inflation}},
  \href{http://dx.doi.org/10.1088/1475-7516/2011/01/014}{\emph{JCAP} {\bfseries
  01} (2011) 014}, [\href{https://arxiv.org/abs/1009.2497}{{\ttfamily
  1009.2497}}].

\bibitem{deRham:2014wfa}
C.~de~Rham and R.~H. Ribeiro, \emph{{Riding on irrelevant operators}},
  \href{http://dx.doi.org/10.1088/1475-7516/2014/11/016}{\emph{JCAP} {\bfseries
  1411} (2014) 016}, [\href{https://arxiv.org/abs/1405.5213}{{\ttfamily
  1405.5213}}].

\bibitem{deRham:2012ew}
C.~de~Rham, G.~Gabadadze, L.~Heisenberg and D.~Pirtskhalava,
  \emph{{Nonrenormalization and naturalness in a class of scalar-tensor
  theories}}, \href{http://dx.doi.org/10.1103/PhysRevD.87.085017}{\emph{Phys.
  Rev. D} {\bfseries 87} (2013) 085017},
  [\href{https://arxiv.org/abs/1212.4128}{{\ttfamily 1212.4128}}].

\bibitem{Noller:2018eht}
J.~Noller and A.~Nicola, \emph{{Radiative stability and observational
  constraints on dark energy and modified gravity}},
  \href{http://dx.doi.org/10.1103/PhysRevD.102.104045}{\emph{Phys. Rev. D}
  {\bfseries 102} (2020) 104045},
  [\href{https://arxiv.org/abs/1811.03082}{{\ttfamily 1811.03082}}].

\bibitem{Heisenberg:2020cyi}
L.~Heisenberg, J.~Noller and J.~Zosso, \emph{{Horndeski under the quantum
  loupe}}, \href{http://dx.doi.org/10.1088/1475-7516/2020/10/010}{\emph{JCAP}
  {\bfseries 10} (2020) 010},
  [\href{https://arxiv.org/abs/2004.11655}{{\ttfamily 2004.11655}}].

\bibitem{Melville:2019wyy}
S.~Melville and J.~Noller, \emph{{Positivity in the Sky: Constraining dark
  energy and modified gravity from the UV}},
  \href{http://dx.doi.org/10.1103/PhysRevD.101.021502}{\emph{Phys. Rev. D}
  {\bfseries 101} (2020) 021502},
  [\href{https://arxiv.org/abs/1904.05874}{{\ttfamily 1904.05874}}].

\bibitem{TheLIGOScientific:2017qsa}
{\scshape LIGO Scientific, Virgo} collaboration, B.~Abbott et~al.,
  \emph{{GW170817: Observation of Gravitational Waves from a Binary Neutron
  Star Inspiral}},
  \href{http://dx.doi.org/10.1103/PhysRevLett.119.161101}{\emph{Phys. Rev.
  Lett.} {\bfseries 119} (2017) 161101},
  [\href{https://arxiv.org/abs/1710.05832}{{\ttfamily 1710.05832}}].

\bibitem{Monitor:2017mdv}
{\scshape LIGO Scientific, Virgo, Fermi-GBM, INTEGRAL} collaboration, B.~P.
  Abbott et~al., \emph{{Gravitational Waves and Gamma-rays from a Binary
  Neutron Star Merger: GW170817 and GRB 170817A}},
  \href{http://dx.doi.org/10.3847/2041-8213/aa920c}{\emph{Astrophys. J.}
  {\bfseries 848} (2017) L13},
  [\href{https://arxiv.org/abs/1710.05834}{{\ttfamily 1710.05834}}].

\bibitem{GBM:2017lvd}
B.~P. Abbott et~al., \emph{{Multi-messenger Observations of a Binary Neutron
  Star Merger}},
  \href{http://dx.doi.org/10.3847/2041-8213/aa91c9}{\emph{Astrophys. J.}
  {\bfseries 848} (2017) L12},
  [\href{https://arxiv.org/abs/1710.05833}{{\ttfamily 1710.05833}}].

\bibitem{Lombriser:2015sxa}
L.~Lombriser and A.~Taylor, \emph{{Breaking a Dark Degeneracy with
  Gravitational Waves}},
  \href{http://dx.doi.org/10.1088/1475-7516/2016/03/031}{\emph{JCAP} {\bfseries
  1603} (2016) 031}, [\href{https://arxiv.org/abs/1509.08458}{{\ttfamily
  1509.08458}}].

\bibitem{Lombriser:2016yzn}
L.~Lombriser and N.~A. Lima, \emph{{Challenges to Self-Acceleration in Modified
  Gravity from Gravitational Waves and Large-Scale Structure}},
  \href{http://dx.doi.org/10.1016/j.physletb.2016.12.048}{\emph{Phys. Lett.}
  {\bfseries B765} (2017) 382--385},
  [\href{https://arxiv.org/abs/1602.07670}{{\ttfamily 1602.07670}}].

\bibitem{Creminelli:2017sry}
P.~Creminelli and F.~Vernizzi, \emph{{Dark Energy after GW170817 and
  GRB170817A}},
  \href{http://dx.doi.org/10.1103/PhysRevLett.119.251302}{\emph{Phys. Rev.
  Lett.} {\bfseries 119} (2017) 251302},
  [\href{https://arxiv.org/abs/1710.05877}{{\ttfamily 1710.05877}}].

\bibitem{Sakstein:2017xjx}
J.~Sakstein and B.~Jain, \emph{{Implications of the Neutron Star Merger
  GW170817 for Cosmological Scalar-Tensor Theories}},
  \href{http://dx.doi.org/10.1103/PhysRevLett.119.251303}{\emph{Phys. Rev.
  Lett.} {\bfseries 119} (2017) 251303},
  [\href{https://arxiv.org/abs/1710.05893}{{\ttfamily 1710.05893}}].

\bibitem{Ezquiaga:2017ekz}
J.~M. Ezquiaga and M.~Zumalac\'arregui, \emph{{Dark Energy After GW170817: Dead
  Ends and the Road Ahead}},
  \href{http://dx.doi.org/10.1103/PhysRevLett.119.251304}{\emph{Phys. Rev.
  Lett.} {\bfseries 119} (2017) 251304},
  [\href{https://arxiv.org/abs/1710.05901}{{\ttfamily 1710.05901}}].

\bibitem{Baker:2017hug}
T.~Baker, E.~Bellini, P.~G. Ferreira, M.~Lagos, J.~Noller and I.~Sawicki,
  \emph{{Strong constraints on cosmological gravity from GW170817 and GRB
  170817A}},
  \href{http://dx.doi.org/10.1103/PhysRevLett.119.251301}{\emph{Phys. Rev.
  Lett.} {\bfseries 119} (2017) 251301},
  [\href{https://arxiv.org/abs/1710.06394}{{\ttfamily 1710.06394}}].

\bibitem{Akrami:2018yjz}
Y.~Akrami, P.~Brax, A.-C. Davis and V.~Vardanyan, \emph{{Neutron star merger
  GW170817 strongly constrains doubly coupled bigravity}},
  \href{http://dx.doi.org/10.1103/PhysRevD.97.124010}{\emph{Phys. Rev.}
  {\bfseries D97} (2018) 124010},
  [\href{https://arxiv.org/abs/1803.09726}{{\ttfamily 1803.09726}}].

\bibitem{Heisenberg:2017qka}
L.~Heisenberg and S.~Tsujikawa, \emph{{Dark energy survivals in massive gravity
  after GW170817: SO(3) invariant}},
  \href{http://dx.doi.org/10.1088/1475-7516/2018/01/044}{\emph{JCAP} {\bfseries
  1801} (2018) 044}, [\href{https://arxiv.org/abs/1711.09430}{{\ttfamily
  1711.09430}}].

\bibitem{BeltranJimenez:2018ymu}
J.~Beltr\`an~Jim\'enez and L.~Heisenberg, \emph{{Non-trivial gravitational
  waves and structure formation phenomenology from dark energy}},
  \href{http://dx.doi.org/10.1088/1475-7516/2018/09/035}{\emph{JCAP} {\bfseries
  1809} (2018) 035}, [\href{https://arxiv.org/abs/1806.01753}{{\ttfamily
  1806.01753}}].

\bibitem{deRham:2018red}
C.~de~Rham and S.~Melville, \emph{{Gravitational Rainbows: LIGO and Dark Energy
  at its Cutoff}},
  \href{http://dx.doi.org/10.1103/PhysRevLett.121.221101}{\emph{Phys. Rev.
  Lett.} {\bfseries 121} (2018) 221101},
  [\href{https://arxiv.org/abs/1806.09417}{{\ttfamily 1806.09417}}].

\bibitem{Pham:1985cr}
T.~Pham and T.~N. Truong, \emph{{Evaluation of the Derivative Quartic Terms of
  the Meson Chiral Lagrangian From Forward Dispersion Relation}},
  \href{http://dx.doi.org/10.1103/PhysRevD.31.3027}{\emph{Phys. Rev. D}
  {\bfseries 31} (1985) 3027}.

\bibitem{Ananthanarayan:1994hf}
B.~Ananthanarayan, D.~Toublan and G.~Wanders, \emph{{Consistency of the chiral
  pion pion scattering amplitudes with axiomatic constraints}},
  \href{http://dx.doi.org/10.1103/PhysRevD.51.1093}{\emph{Phys. Rev. D}
  {\bfseries 51} (1995) 1093--1100},
  [\href{https://arxiv.org/abs/hep-ph/9410302}{{\ttfamily hep-ph/9410302}}].

\bibitem{Pennington:1994kc}
M.~Pennington and J.~Portoles, \emph{{The Chiral Lagrangian parameters, l1, l2,
  are determined by the rho resonance}},
  \href{http://dx.doi.org/10.1016/0370-2693(94)01551-M}{\emph{Phys. Lett. B}
  {\bfseries 344} (1995) 399--406},
  [\href{https://arxiv.org/abs/hep-ph/9409426}{{\ttfamily hep-ph/9409426}}].

\bibitem{Comellas:1995hq}
J.~Comellas, J.~I. Latorre and J.~Taron, \emph{{Constraints on chiral
  perturbation theory parameters from QCD inequalities}},
  \href{http://dx.doi.org/10.1016/0370-2693(95)01110-C}{\emph{Phys. Lett. B}
  {\bfseries 360} (1995) 109--116},
  [\href{https://arxiv.org/abs/hep-ph/9507258}{{\ttfamily hep-ph/9507258}}].

\bibitem{Dvali:2012zc}
G.~Dvali, A.~Franca and C.~Gomez, \emph{{Road Signs for UV-Completion}},
  \href{https://arxiv.org/abs/1204.6388}{{\ttfamily 1204.6388}}.

\bibitem{Bellazzini:2020cot}
B.~Bellazzini, J.~Elias~Mir\'o, R.~Rattazzi, M.~Riembau and F.~Riva,
  \emph{{Positive Moments for Scattering Amplitudes}},
  \href{https://arxiv.org/abs/2011.00037}{{\ttfamily 2011.00037}}.

\bibitem{Tolley:2020gtv}
A.~J. Tolley, Z.-Y. Wang and S.-Y. Zhou, \emph{{New positivity bounds from full
  crossing symmetry}},  \href{https://arxiv.org/abs/2011.02400}{{\ttfamily
  2011.02400}}.

\bibitem{Caron-Huot:2020cmc}
S.~Caron-Huot and V.~Van~Duong, \emph{{Extremal Effective Field Theories}},
  \href{https://arxiv.org/abs/2011.02957}{{\ttfamily 2011.02957}}.

\bibitem{Sinha:2020win}
A.~Sinha and A.~Zahed, \emph{{Crossing Symmetric Dispersion Relations in
  QFTs}},  \href{https://arxiv.org/abs/2012.04877}{{\ttfamily 2012.04877}}.

\bibitem{Alberte:2020bdz}
L.~Alberte, C.~de~Rham, S.~Jaitly and A.~J. Tolley, \emph{{QED positivity
  bounds}},  \href{https://arxiv.org/abs/2012.05798}{{\ttfamily 2012.05798}}.

\bibitem{Caron-Huot:2021rmr}
S.~Caron-Huot, D.~Mazac, L.~Rastelli and D.~Simmons-Duffin, \emph{{Sharp
  Boundaries for the Swampland}},
  \href{https://arxiv.org/abs/2102.08951}{{\ttfamily 2102.08951}}.

\bibitem{Grall:2021xxm}
T.~Grall and S.~Melville, \emph{{Positivity Bounds without Boosts}},
  \href{https://arxiv.org/abs/2102.05683}{{\ttfamily 2102.05683}}.

\bibitem{Vecchi:2007na}
L.~Vecchi, \emph{{Causal versus analytic constraints on anomalous quartic gauge
  couplings}},
  \href{http://dx.doi.org/10.1088/1126-6708/2007/11/054}{\emph{JHEP} {\bfseries
  11} (2007) 054}, [\href{https://arxiv.org/abs/0704.1900}{{\ttfamily
  0704.1900}}].

\bibitem{Manohar:2008tc}
A.~V. Manohar and V.~Mateu, \emph{{Dispersion Relation Bounds for pi pi
  Scattering}}, \href{http://dx.doi.org/10.1103/PhysRevD.77.094019}{\emph{Phys.
  Rev. D} {\bfseries 77} (2008) 094019},
  [\href{https://arxiv.org/abs/0801.3222}{{\ttfamily 0801.3222}}].

\bibitem{Salvatelli:2016mgy}
V.~Salvatelli, F.~Piazza and C.~Marinoni, \emph{{Constraints on modified
  gravity from Planck 2015: when the health of your theory makes the
  difference}},
  \href{http://dx.doi.org/10.1088/1475-7516/2016/09/027}{\emph{JCAP} {\bfseries
  09} (2016) 027}, [\href{https://arxiv.org/abs/1602.08283}{{\ttfamily
  1602.08283}}].

\bibitem{Pujolas:2011he}
O.~Pujolas, I.~Sawicki and A.~Vikman, \emph{{The Imperfect Fluid behind Kinetic
  Gravity Braiding}},
  \href{http://dx.doi.org/10.1007/JHEP11(2011)156}{\emph{JHEP} {\bfseries 11}
  (2011) 156}, [\href{https://arxiv.org/abs/1103.5360}{{\ttfamily 1103.5360}}].

\bibitem{Barreira:2014jha}
A.~Barreira, B.~Li, C.~Baugh and S.~Pascoli, \emph{{The observational status of
  Galileon gravity after Planck}},
  \href{http://dx.doi.org/10.1088/1475-7516/2014/08/059}{\emph{JCAP} {\bfseries
  1408} (2014) 059}, [\href{https://arxiv.org/abs/1406.0485}{{\ttfamily
  1406.0485}}].

\bibitem{Planck-Collaboration:2016af}
{Planck Collaboration}, \emph{{Planck 2015 results. XI. CMB power spectra,
  likelihoods, and robustness of parameters}},
  \href{http://dx.doi.org/10.1051/0004-6361/201526926}{\emph{Astronomy and
  Astrophysics} {\bfseries 594} (Sept., 2016) A11},
  [\href{https://arxiv.org/abs/1507.02704}{{\ttfamily 1507.02704}}].

\bibitem{Planck-Collaboration:2016aa}
{Planck Collaboration}, \emph{{Planck 2015 results. XV. Gravitational
  lensing}},
  \href{http://dx.doi.org/10.1051/0004-6361/201525941}{\emph{Astronomy and
  Astrophysics} {\bfseries 594} (Sept., 2016) A15},
  [\href{https://arxiv.org/abs/1502.01591}{{\ttfamily 1502.01591}}].

\bibitem{Planck-Collaboration:2016ae}
{Planck Collaboration}, \emph{{Planck 2015 results. XIII. Cosmological
  parameters}},
  \href{http://dx.doi.org/10.1051/0004-6361/201525830}{\emph{Astronomy and
  Astrophysics} {\bfseries 594} (Sept., 2016) A13},
  [\href{https://arxiv.org/abs/1502.01589}{{\ttfamily 1502.01589}}].

\bibitem{Anderson:2014}
L.~{Anderson et al.}, \emph{{The clustering of galaxies in the SDSS-III Baryon
  Oscillation Spectroscopic Survey: baryon acoustic oscillations in the Data
  Releases 10 and 11 Galaxy samples}},
  \href{http://dx.doi.org/10.1093/mnras/stu523}{\emph{\mnras} {\bfseries 441}
  (June, 2014) 24--62}, [\href{https://arxiv.org/abs/1312.4877}{{\ttfamily
  1312.4877}}].

\bibitem{Ross:2015}
A.~J. {Ross}, L.~{Samushia}, C.~{Howlett}, W.~J. {Percival}, A.~{Burden} and
  M.~{Manera}, \emph{{The clustering of the SDSS DR7 main Galaxy sample - I. A
  4 per cent distance measure at z = 0.15}},
  \href{http://dx.doi.org/10.1093/mnras/stv154}{\emph{\mnras} {\bfseries 449}
  (May, 2015) 835--847}, [\href{https://arxiv.org/abs/1409.3242}{{\ttfamily
  1409.3242}}].

\bibitem{Tegmark:2006}
M.~{Tegmark et al.}, \emph{{Cosmological constraints from the SDSS luminous red
  galaxies}}, \href{http://dx.doi.org/10.1103/PhysRevD.74.123507}{\emph{\prd}
  {\bfseries 74} (Dec., 2006) 123507},
  [\href{https://arxiv.org/abs/astro-ph/0608632}{{\ttfamily
  astro-ph/0608632}}].

\bibitem{Beutler:2012}
F.~{Beutler}, C.~{Blake}, M.~{Colless}, D.~H. {Jones}, L.~{Staveley-Smith},
  G.~B. {Poole} et~al., \emph{{The 6dF Galaxy Survey: $z \approx 0$
  measurements of the growth rate and {$\sigma$}$_{8}$}},
  \href{http://dx.doi.org/10.1111/j.1365-2966.2012.21136.x}{\emph{\mnras}
  {\bfseries 423} (July, 2012) 3430--3444},
  [\href{https://arxiv.org/abs/1204.4725}{{\ttfamily 1204.4725}}].

\bibitem{Samushia:2014}
L.~{Samushia et al.}, \emph{{The clustering of galaxies in the SDSS-III Baryon
  Oscillation Spectroscopic Survey: measuring growth rate and geometry with
  anisotropic clustering}},
  \href{http://dx.doi.org/10.1093/mnras/stu197}{\emph{\mnras} {\bfseries 439}
  (Apr., 2014) 3504--3519}, [\href{https://arxiv.org/abs/1312.4899}{{\ttfamily
  1312.4899}}].

\bibitem{Kennedy:2020ehn}
J.~Kennedy and L.~Lombriser, \emph{{Positivity bounds on reconstructed
  Horndeski models}},
  \href{http://dx.doi.org/10.1103/PhysRevD.102.044062}{\emph{Phys. Rev. D}
  {\bfseries 102} (2020) 044062},
  [\href{https://arxiv.org/abs/2003.05318}{{\ttfamily 2003.05318}}].

\bibitem{Burrage:2020jkj}
C.~Burrage and J.~Dombrowski, \emph{{Constraining the cosmological evolution of
  scalar-tensor theories with local measurements of the time variation of G}},
  \href{http://dx.doi.org/10.1088/1475-7516/2020/07/060}{\emph{JCAP} {\bfseries
  07} (2020) 060}, [\href{https://arxiv.org/abs/2004.14260}{{\ttfamily
  2004.14260}}].

\bibitem{Kobayashi:2011nu}
T.~Kobayashi, M.~Yamaguchi and J.~Yokoyama, \emph{{Generalized G-inflation:
  Inflation with the most general second-order field equations}},
  \href{http://dx.doi.org/10.1143/PTP.126.511}{\emph{Prog. Theor. Phys.}
  {\bfseries 126} (2011) 511--529},
  [\href{https://arxiv.org/abs/1105.5723}{{\ttfamily 1105.5723}}].

\bibitem{Creminelli:2019kjy}
P.~Creminelli, G.~Tambalo, F.~Vernizzi and V.~Yingcharoenrat,
  \emph{{Dark-Energy Instabilities induced by Gravitational Waves}},
  \href{http://dx.doi.org/10.1088/1475-7516/2020/05/002}{\emph{JCAP} {\bfseries
  05} (2020) 002}, [\href{https://arxiv.org/abs/1910.14035}{{\ttfamily
  1910.14035}}].

\bibitem{Milonni}
P.~Milonni, \emph{{Fast Light, Slow Light and Left-Handed Light, (Series in
  Optics and Optoelectronics)}}, {\emph{{Fast Light, Slow Light and Left-Handed
  Light\!\!}} (2004) 262 pages}, [\href{https://arxiv.org/abs/(Taylor \&
  Francis)}{{\ttfamily (Taylor \& Francis)}}].

\bibitem{Brillouin}
L.~Brillouin, \emph{{Wave Propagation and Group Velocity (Series in Pure \&
  Applied Physics) }}, {\emph{{Wave Propagation and Group Velocity\!\!}} (1960)
  154 pages}, [\href{https://arxiv.org/abs/(Academic Press)}{{\ttfamily
  (Academic Press)}}].

\bibitem{deRham:2014zqa}
C.~de~Rham, \emph{{Massive Gravity}},
  \href{http://dx.doi.org/10.12942/lrr-2014-7}{\emph{Living Rev. Rel.}
  {\bfseries 17} (2014) 7}, [\href{https://arxiv.org/abs/1401.4173}{{\ttfamily
  1401.4173}}].

\bibitem{Reall:2021voz}
H.~S. Reall, \emph{{Causality in gravitational theories with second order
  equations of motion}},  \href{https://arxiv.org/abs/2101.11623}{{\ttfamily
  2101.11623}}.

\bibitem{Blas:2011rf}
D.~Blas, J.~Lesgourgues and T.~Tram, \emph{{The Cosmic Linear Anisotropy
  Solving System (CLASS) II: Approximation schemes}},
  \href{http://dx.doi.org/10.1088/1475-7516/2011/07/034}{\emph{JCAP} {\bfseries
  1107} (2011) 034}, [\href{https://arxiv.org/abs/1104.2933}{{\ttfamily
  1104.2933}}].

\bibitem{corner}
D.~Foreman-Mackey, \emph{corner.py: Scatterplot matrices in python},
  \href{http://dx.doi.org/10.21105/joss.00024}{\emph{The Journal of Open Source
  Software} {\bfseries 24} (2016) }.

\bibitem{Bellini:2019syt}
E.~Bellini, I.~Sawicki and M.~Zumalac\'arregui, \emph{{hi\_class: Background
  Evolution, Initial Conditions and Approximation Schemes}},
  \href{http://dx.doi.org/10.1088/1475-7516/2020/02/008}{\emph{JCAP} {\bfseries
  02} (2020) 008}, [\href{https://arxiv.org/abs/1909.01828}{{\ttfamily
  1909.01828}}].

\bibitem{Audren:2012wb}
B.~Audren, J.~Lesgourgues, K.~Benabed and S.~Prunet, \emph{{Conservative
  Constraints on Early Cosmology: an illustration of the Monte Python
  cosmological parameter inference code}},
  \href{http://dx.doi.org/10.1088/1475-7516/2013/02/001}{\emph{JCAP} {\bfseries
  1302} (2013) 001}, [\href{https://arxiv.org/abs/1210.7183}{{\ttfamily
  1210.7183}}].

\bibitem{Brinckmann:2018cvx}
T.~Brinckmann and J.~Lesgourgues, \emph{{MontePython 3: boosted MCMC sampler
  and other features}},  \href{https://arxiv.org/abs/1804.07261}{{\ttfamily
  1804.07261}}.

\bibitem{xAct}
J.~M. Mart\'in-Garc\'ia, \emph{{xAct 2002-2014}}, {\emph{http://www.xact.es/}
  }.

\bibitem{Aragone:1979hx}
C.~Aragone and S.~Deser, \emph{{Consistency Problems of Hypergravity}},
  \href{http://dx.doi.org/10.1016/0370-2693(79)90808-6}{\emph{Phys. Lett. B}
  {\bfseries 86} (1979) 161--163}.

\bibitem{Porrati:1993in}
M.~Porrati, \emph{{Massive spin 5/2 fields coupled to gravity: Tree level
  unitarity versus the equivalence principle}},
  \href{http://dx.doi.org/10.1016/0370-2693(93)91403-A}{\emph{Phys. Lett. B}
  {\bfseries 304} (1993) 77--80},
  [\href{https://arxiv.org/abs/gr-qc/9301012}{{\ttfamily gr-qc/9301012}}].

\bibitem{Cucchieri:1994tx}
A.~Cucchieri, M.~Porrati and S.~Deser, \emph{{Tree level unitarity constraints
  on the gravitational couplings of higher spin massive fields}},
  \href{http://dx.doi.org/10.1103/PhysRevD.51.4543}{\emph{Phys. Rev. D}
  {\bfseries 51} (1995) 4543--4549},
  [\href{https://arxiv.org/abs/hep-th/9408073}{{\ttfamily hep-th/9408073}}].

\end{thebibliography}\endgroup
\end{document}